\documentclass[journal,10pt]{IEEEtran}
\usepackage{graphicx}
\usepackage{caption}
\usepackage{subcaption}
\usepackage{mathtools}
\usepackage{bm, amsmath, amssymb}
\usepackage{amsmath}
\usepackage{amsfonts}
\usepackage {amssymb,amsmath}
\usepackage{cite}
\usepackage{epstopdf}
\usepackage{microtype}
\DeclareGraphicsExtensions{.pdf,.png,.jpg}

\newcommand {\Define} {\stackrel {\Delta} {=}  }
\newcommand{\mya}{\mathrel{\overset{\makebox[0pt]{{\tiny(a)}}}{=}}}
\newcommand{\myb}{\mathrel{\overset{\makebox[0pt]{{\tiny(b)}}}{=}}}
\newcommand{\myc}{\mathrel{\overset{\makebox[0pt]{{\tiny(c)}}}{=}}}
\newcommand{\myd}{\mathrel{\overset{\makebox[0pt]{{\tiny(d)}}}{=}}}

\newcommand{\myapproxa}{\mathrel{\overset{\makebox[0pt]{{\tiny(a)}}}{\approx}}}

\newtheorem{theorem}{Theorem}

\newtheorem{lemma}{Lemma}
\newtheorem{result}{Result}
\usepackage{cite}

\begin{document}
\title{Derivation of OTFS Modulation from First Principles}
\author{\IEEEauthorblockN{Saif Khan Mohammed, Senior Member, IEEE}
\IEEEauthorblockA{ \thanks{Saif Khan Mohammed is with the Department of Electrical Engineering, Indian Institute of Technology Delhi, New Delhi, India. Email: saifkmohammed@gmail.com. This work is supported by the Prof. Kishan Gupta and Pramila Gupta Chair at IIT Delhi.}}
}
\maketitle

\vspace{-10mm}
\begin{abstract}
Orthogonal Time Frequency Space (OTFS) modulation has been recently proposed to be
robust to channel induced Doppler shift in high mobility wireless communication systems. However, to the best of our knowledge,
none of the prior works on OTFS have derived it from first principles.  
In this paper, using the ZAK representation of time-domain (TD) signals, we rigorously derive an orthonormal basis of approximately time and bandwidth limited signals which are also localized in the delay-Doppler (DD) domain. We then consider DD domain modulation based on this orthonormal basis, and derive OTFS modulation.
To the best of our knowledge, this is the first paper to rigorously derive OTFS modulation from first principles.
We show that irrespective of the amount of Doppler shift, the received DD domain basis signals are localized
in a small interval of size roughly equal to the inverse time duration along the Doppler domain and of size roughly equal to the inverse bandwidth along the delay domain (time duration refers to the
length of the time-interval where the TD transmit signal has been limited). 
With sufficiently large time duration and bandwidth, there is little interference between information symbols modulated on different basis signals, which allows for
joint DD domain equalization of all information symbols. This explains the inherent robustness of DD domain modulation to channel induced Doppler shift when compared with Orthogonal Frequency Division Multiplexing (OFDM).
The degree of localization of the DD domain basis signals is inversely related to the time duration of the transmit signal, which explains the trade-off between
robustness to Doppler shift and latency.                    
\end{abstract}

\begin{IEEEkeywords}
	Orthonormal Basis, Delay-Doppler, ZAK Representation, OTFS, Doppler Shift.
\end{IEEEkeywords}
\section{Introduction}
Next generation wireless communication systems are expected to support reliable and high data rate communication
even at very high mobile speed \cite{IMT2020}. 
However, the modulation waveform used in Fifth Generation (5G) communication systems is based
on Orthogonal Frequency Division Multiplexing (OFDM) for which communication reliability and data rate is known
to degrade in high mobility scenarios \cite{NR5G}.  
Recently, Orthogonal Time Frequency Space (OTFS) modulation has been proposed to be robust
to channel induced Doppler shift when compared to OFDM \cite{HadaniOTFS1, HadaniOTFS2, HadaniOTFS3}.
In OTFS modulation, information is embedded in the delay-Doppler (DD) domain. The information bearing
DD signal is then converted to a time-domain (TD) transmit signal.
At the receiver, the received TD signal is converted to a DD domain signal from which the information symbols are decoded,
i.e., modulation, demodulation and channel estimation are all performed in the DD domain \cite{channel,mimootfs,MCMC,Estimation}.

To the best of our knowledge, none of the prior works have rigorously derived OTFS modulation from first principles.
In the absence of a rigorous mathematical derivation of OTFS modulation and its basis waveforms,
it is difficult for communication engineers to fully understand the robustness of OTFS modulation to Doppler shift.
A deeper understanding of DD domain modulation and basis waveforms is required to design modulation and demodulation
methods which are robust to channel induced Doppler shift in very high mobility scenarios (e.g., high speed train, air-to-ground communication).    
Therefore, in this paper, using the ZAK representation/transform\footnote{\footnotesize{ZAK representation is named after its inventor, J. Zak \cite{Zak1}. In this current paper, ``ZAK representation" refers to the delay-Doppler domain representation of TD signals, as defined in \cite{Janssen}.}} of TD signals, we rigorously derive OTFS modulation from first principles. 
The novel contributions of this paper are:
\begin{itemize}
\item In Section \ref{seczak}, using the ZAK representation of TD signals, we derive an expression for TD signals which are neither time-limited nor bandwidth limited, but
which are perfectly localized (i.e., Dirac-delta impulse) in the DD domain. In Lemma \ref{plem1st} we show that these TD signals are an impulse train which is time-shifted and multiplied by a complex exponential. In Theorem \ref{thmbasis}
we show that these signals form a basis for all TD signals. 
\item The TD basis signals derived in Section  \ref{seczak} are neither time-limited nor bandwidth limited. Therefore, in Section \ref{secbw},
by approximately limiting the TD basis signals of Section \ref{seczak} along time and frequency domains, we obtain the expression for TD signals which 
are approximately time and bandwidth limited. We then derive the ZAK representation of these TD signals in Theorem \ref{thm21}.
We show that due to time and bandwidth limitation, the corresponding DD domain signal is no more perfectly localized at a point but is instead spread over
an interval whose size along the delay and Doppler domains is roughly equal to the inverse bandwidth and inverse time duration respectively.
\item Further, in Theorem \ref{lem18} in Section \ref{secbw}, for a given $(T , \Delta f)$, $\Delta f = 1/T$, and positive integers $M, N$, we derive a basis of orthonormal signals which are approximately time-limited to $NT$ seconds
and bandwidth limited to $M \Delta f$ Hz, and are
also localized in an interval of size inverse-bandwidth and inverse-time duration along the delay and Doppler domain respectively. The dimensionality of this basis
is equal to the time-bandwidth product $NT \times M \Delta f = MN$. Our derivation therefore shows the important result that
the additional constraint of DD domain localization does not reduce the dimensionality of approximately time and bandwidth limited signals.  
\item Using the orthonormal basis derived in Section \ref{secbw}, in Section \ref{propmodulation} we consider DD domain modulation, where DD domain information symbols
linearly modulate the orthonormal DD domain basis signals derived in Section \ref{secbw}. 
\item In Theorem \ref{thmotfsmod} in Section \ref{deriveOTFS}, we derive OTFS modulation from the DD domain modulation derived in Section \ref{propmodulation}.   
To the best of our knowledge, this paper is the first to rigorously derive OTFS modulation from first principles.
\item In Section \ref{Zakrecv} we derive an expression for the spectral efficiency (SE) achieved by the DD domain modulation derived in Section \ref{propmodulation}.
\item In Section \ref{secbetter}, we study the localization of the received basis signals in the DD domain. We show that the energy transmitted on a particular DD domain basis signal interferes with only a small fraction of
the other $(MN -1)$ basis signals when compared to the fraction of interfered sub-carriers in OFDM. With increasing Doppler shift, the variation in the fraction of interfered DD domain basis signals is
much smaller than the variation in the fraction of interfered sub-carriers in OFDM. This explains the inherent robustness of DD domain modulation to channel induced Doppler shift.
\item It is also observed that for a given $M$ (i.e., given bandwidth $M \Delta f$), the fraction of interfered DD domain basis signals decreases with increasing $N$ which makes it easier
to perform joint equalization of all $MN$ information symbols in the DD domain. However, with increasing $N$ the time duration $NT$ increases, which increases latency.    
\end{itemize}                       

{\em Notations:} 
The continuous-time Dirac-delta signal with impulse at $t=0$ is denoted by $\delta(t)$. The discrete-time
impulse signal is denoted by $\delta[ k], k \in {\mathbb Z}$, where $\delta[k] = 1$ for $k=0$ and is zero otherwise.
For any matrix ${\bf A}$, $\left\vert {\bf A} \right\vert$ denotes the determinant of ${\bf A}$. Also, $A[p,q]$ denotes the element in the $p$-th row and $q$-th column of matrix ${\bf A}$.
The zero mean circular symmetric complex Gaussian distribution with variance $\sigma^2$ is denoted by ${\mathcal C}{\mathcal N}(0, \sigma^2)$.
The conjugate of a complex number $z \in {\mathbb C}$ is denoted
by $z^*$. The real part of a complex number $z$ is denoted by $Re(z)$. For any real number $x$, $\lfloor x \rfloor$ denotes the greatest integer smaller than or equal to $x$.
For any integer $M$ and real number $x$, $[ x ]_{_M}$ denotes the smallest unique non-negative real number such that $\left(x -  [ x ]_{_M} \right)$
is an integer multiple of $M$. The symbol $\%$ denote percent, e.g. $12.5 \%$ is $0.125$.
For any set ${\mathcal A}$, $\left\vert {\mathcal A} \right\vert$ denotes its cardinality.
The abbreviation R.H.S. stands for ``right hand side" and w.r.t. stands for ``with respect to".
For any real $x$, $\mbox{\small{sinc}}(x) \Define \frac{\sin(\pi x)}{\pi x}$.
For any two sets ${\mathcal A}$ and ${\mathcal B}$, ${\mathcal A} \subseteq {\mathcal B}$
means that ${\mathcal A}$ is a subset of ${\mathcal B}$. 

\section{The ZAK Representation of Time-Domain (TD) Signals}
\label{seczak}
Let $x(t)$ be a complex time-continuous signal. For any $T > 0$ we define the ZAK representation of $x(t)$ by the two-dimensional signal \cite{Janssen}
\begin{eqnarray}
\label{zdef1}
{\mathcal Z}_x(\tau, \nu) & \Define & \sqrt{T} \, \sum\limits_{n=-\infty}^{\infty} \, x(\tau + nT) \, e^{-j 2 \pi n \nu T } \,,\, \nonumber \\
& & -\infty < \tau < \infty \,,\, -\infty < \nu < \infty. 
\end{eqnarray}
The following result from \cite{Saif2020} shows that
a channel induced time shift (due to path delay) and frequency shift (due to Doppler) to a TD signal $x(t)$
corresponds to simple shifts along the $\tau-$ and $\nu-$ domains in its ZAK representation.   
Therefore, subsequently we refer to the $\tau-$ and $\nu-$ domains as
the ``delay" and ``Doppler" domain respectively, i.e., jointly we refer to them as the delay-Doppler (DD) domain. 
\begin{result}\mbox{[see Theorem $1$ in \cite{Saif2020}]}
\label{zlem52}
Let there be only one channel path with
a delay of $\tau_0$ and a Doppler shift of $\nu_0$. With $x(t)$ as the transmit signal, the ZAK representation of the noise-free received signal
$r(t) = x(t - \tau_0) e^{j 2 \pi \nu_0 (t - \tau_0)}$ is given by
\begin{eqnarray}
\label{zakmod2}
{\mathcal Z}_r(\tau, \nu) & = & e^{j 2 \pi \nu_0 (\tau - \tau_0)} {\mathcal Z}_x(\tau - \tau_0 , \nu - \nu_0)
\end{eqnarray}
i.e., delay and Doppler shift in TD results in a shift of $\tau_0$ and $\nu_0$ along the $\tau-$ and $\nu-$ domains respectively. 
\end{result}
\begin{IEEEproof}
See proof of Theorem $1$ in \cite{Saif2020}.
\end{IEEEproof}
 
In the following we present important results on ZAK representation which will be useful later. These
results are available in \cite{Janssen} for normalized $T = \Delta f = 1$. Here we present these
results for a general $T$ and $\Delta f = 1/T$, and for the general audience we also provide much simpler and detailed step by step proof of these
results in the appendix. These results from \cite{Janssen} have been mentioned as ``Result", whereas our original/novel results stated and proved in this current paper have been referred to
as ``Lemma" or ``Theorem".
 
The following result states that the ZAK representation of a TD signal is {\em quasi-periodic} along the delay and Doppler domain.  
\begin{result}\mbox{[see $(2.20)$ and $(2.21)$ in \cite{Janssen}]}
\label{zlem1}
For any $x(t)$, the corresponding ZAK representation ${\mathcal Z}_x(\tau,\nu)$ is periodic along the Doppler domain with a period of $\Delta f = 1/T$ and is quasi-periodic
along the delay domain with a period $T$, i.e.,
\begin{eqnarray}
\label{zeqn2}
{\mathcal Z}_x(\tau + T, \nu) & = & e^{j 2 \pi \nu T} \, {\mathcal Z}_x(\tau,\nu) \,,\, \nonumber \\
{\mathcal Z}_x \left(\tau,\nu + \Delta f \right) & = & {\mathcal Z}_x(\tau,\nu).
\end{eqnarray}
\end{result}
\begin{IEEEproof}
See Appendix \ref{prf_zlem1}.
\end{IEEEproof}
From (\ref{zeqn2}) it follows that for any integer $n$, ${\mathcal Z}_x(\tau + nT, \nu) = e^{j 2 \pi \nu T} \, {\mathcal Z}_x(\tau + (n-1)T,\nu)$,
repeated use of which gives
\begin{eqnarray}
\label{zquasieqn1}
{\mathcal Z}_x(\tau + nT, \nu) = e^{j 2 \pi \nu n T} \, {\mathcal Z}_x(\tau ,\nu), \,\,\, n \in {\mathbb Z}.
\end{eqnarray}
Similarly, from (\ref{zeqn2}) it also follows that
\begin{eqnarray}
\label{zquasieqn2}
{\mathcal Z}_x(\tau, \nu + m \Delta f) = {\mathcal Z}_x(\tau ,\nu), \,\,\, m \in {\mathbb Z}.
\end{eqnarray}

Conversely, it is also true that if a DD domain signal ${\mathcal Z}_x(\tau,\nu)$ satisfies the quasi-periodicity conditions
in (\ref{zeqn2}), then there exists a unique TD signal $x(t)$ whose ZAK representation is ${\mathcal Z}_x(\tau,\nu)$ (see $(2.35)$ and $(2.36)$ in \cite{Janssen}).  

The TD signal $x(t)$ and its Fourier transform ${\mathcal F}_x(f) = \int\limits_{-\infty}^{\infty} x(t) e^{-j 2 \pi f t} \, dt$ can be obtained from the ZAK representation
${\mathcal Z}_x(\tau,\nu)$ as stated in the following result.
\begin{result}\mbox{[see $(2.29)$ and $(2.30)$ in \cite{Janssen}]}
\label{zlem3}
The TD signal $x(t)$ can be recovered from its ZAK representation by
\begin{eqnarray}
\label{zeqn6}
x(t) & = & \sqrt{T} \int\limits_{0}^{\Delta f} \, {\mathcal Z}_x(t, \nu) \, d\nu
\end{eqnarray}
and the Fourier transform of $x(t)$ is given by
\begin{eqnarray}
\label{zeqn7}
{\mathcal F}_x(f) & = & \frac{1}{\sqrt{T}} \int\limits_{0}^T {\mathcal Z}_x(\tau, f) \, e^{-j 2 \pi f \tau} \, d\tau.
\end{eqnarray}
\end{result}
\begin{IEEEproof}
See Appendix \ref{prf_zlem3}.
\end{IEEEproof}

\begin{figure}[t]
\vspace{-0.7 cm}
\hspace{-0.2 in}
\centering
\includegraphics[width= 3.6 in, height= 2.9 in]{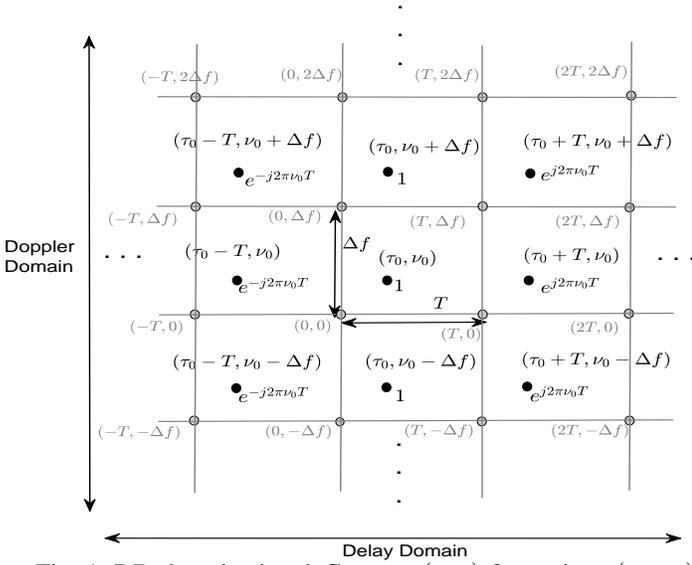}
\vspace{-0.2 cm}
\caption{DD domain signal ${\mathcal Z}_{(p, \tau_0, \nu_0)}(\tau, \nu)$ for a given $(\tau_0, \nu_0)$.} 
\vspace{-0.2cm}
\label{fig1}
\end{figure}

We know that it is not possible to simultaneously localize a signal in the time as well as in the frequency domain,
i.e., there exists no TD signal $x(t)$ which is zero outside some interval $[T_1 \,,\, T_2]$ and whose Fourier transform
${\mathcal F}_x(f)$ is also zero outside some interval $[F_1 \,,\, F_2]$, where $T_1, T_2, F_1, F_2$ are all finite. 
However, there exists TD signals which are simultaneously localized in the delay as well as the Doppler
domain. A DD domain signal localized at $\tau = \tau_0$ along the delay domain and at $\nu = \nu_0$ along the Doppler domain ($0 \leq \tau_0 < T$, $0 \leq \nu_0 < \Delta f$) is given by
\begin{eqnarray}
\label{lceqn1}
{\mathcal Z}_{(p, \tau_0, \nu_0)}(\tau, \nu) & \hspace{-2mm}  \Define & \hspace{-2mm}
\sum\limits_{m=-\infty}^{\infty} \sum\limits_{n = -\infty}^{\infty}  {\Big (} e^{j 2 \pi \nu_0 n T} \delta(\tau - \tau_0 - nT)  \nonumber \\
& &  \hspace{24mm} \delta(\nu - \nu_0 - m \Delta f)  {\Big )}
\end{eqnarray} 
where $\delta(\tau)$ and $\delta(\nu)$ are the Dirac-delta impulse signal along the delay and Doppler domains respectively.
In Fig.~\ref{fig1} we illustrate the signal ${\mathcal Z}_{(p, \tau_0, \nu_0)}(\tau, \nu)$ in the DD domain. In Fig.~\ref{fig1}, the location of the impulses (see R.H.S. in (\ref{lceqn1}))
are denoted by dark dots. The complex value and co-ordinate of each DD domain impulse is mentioned next to it.
Only a portion of the DD domain is illustrated as the delay and Doppler domains extend infinitely in both directions.

\begin{figure}[t]
\vspace{-0.3 cm}
\hspace{-0.3 in}
\centering
\includegraphics[width= 3.85 in, height= 2.9 in]{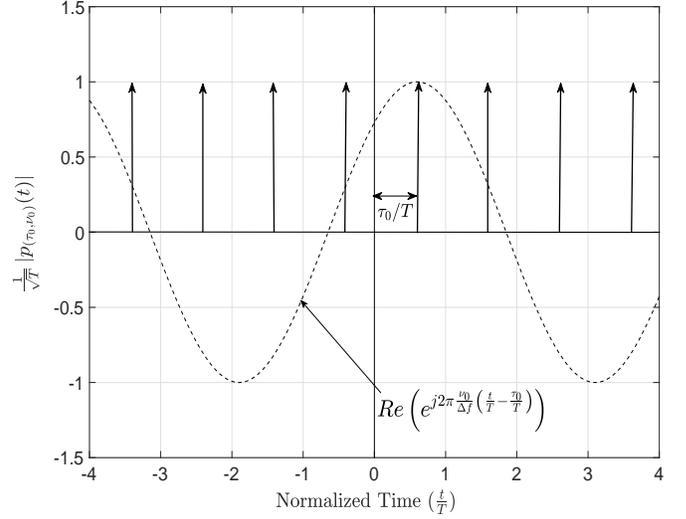}
\vspace{-0.2 cm}
\caption{$\frac{1}{\sqrt{T}} \left\vert  p_{(\tau_0, \nu_0)}(t)  \right\vert$ vs. $t/T$ for $\tau_0 = 0.6 \, T$, $\nu_0 = 0.2 \, \Delta f$.} 
\vspace{-0.2cm}
\label{fig2}
\end{figure}
 
This DD domain signal ${\mathcal Z}_{(p, \tau_0, \nu_0)}(\tau, \nu)$ satisfies the quasi-periodicity conditions in (\ref{zeqn2}) and therefore
the corresponding TD signal $p_{(\tau_0,\nu_0)}(t)$ is given by the following Lemma.
\begin{lemma}
\label{plem1st}
The TD signal $p_{(\tau_0, \nu_0)}(t)$ having ZAK representation ${\mathcal Z}_{(p, \tau_0, \nu_0)}(\tau, \nu)$ in (\ref{lceqn1}) is given by
\begin{eqnarray}
\label{plemeqn2}
p_{(\tau_0, \nu_0)}(t) & = & \sqrt{T} \sum\limits_{n=-\infty}^{\infty} e^{j 2 \pi \nu_0 n T } \, \delta(t  - \tau_0 - nT).
\end{eqnarray}
\end{lemma} 
\begin{IEEEproof}
See Appendix \ref{prf_plem1st}.
\end{IEEEproof}
Therefore, for a given $(\tau_0, \nu_0)$, the TD signal $p_{(\tau_0, \nu_0)}(t)$ is essentially an impulse train multiplied
by the complex exponential $e^{j 2 \pi \nu_0 (t - \tau_0) }$. The impulses are spaced $T$ seconds apart and the
impulse in the time-interval $[0 \,,\, T)$ is located at $t = \tau_0$ (see Fig.~\ref{fig2}, where we have plotted $\frac{1}{\sqrt{T}} \left\vert p_{(\tau_0, \nu_0)}(t) \right\vert$ vs. $t/T$ for
$\tau_0 = 0.6 T, \nu_0 = 0.2 \Delta f$).
Further, we note that for any $(\tau_0, \nu_0), 0 \leq \tau_0 < T, 0 \leq \nu_0 < \Delta f$, $p_{(\tau_0, \nu_0)}(t)$ can be obtained
from the TD signal $p_{(0, 0)}(t)$ by firstly multiplying $p_{(0, 0)}(t)$ by $e^{j 2 \pi \nu_0 t}$ (equivalent to a shift by $\nu_0$ along the Doppler domain)
and then delaying this product signal by $\tau_0$ (equivalent to a shift by $\tau_0$ along the delay domain).
    
The next theorem states that the TD signals $p_{(\tau_0, \nu_0)}(t) \,,\, 0 \leq \tau_0 < T \,,\, 0 \leq \nu_0 < \Delta f$, form
a basis for the space of TD signals.
\begin{theorem}
\label{thmbasis}
Any TD signal $x(t)$ can be expressed in terms of the basis signals $p_{(\tau_0, \nu_0)}(t) \,,\, 0 \leq \tau_0 < T \,,\, 0 \leq \nu_0 < \Delta f$, i.e.
\begin{eqnarray}
\label{thmbasiseqn1}
x(t) & = & \int_{0}^T \int_{0}^{\Delta f}  \hspace{-2mm} c_x(\tau_0, \nu_0) \, p_{(\tau_0, \nu_0)}(t) \, d\tau_0 \, d\nu_0, \nonumber \\
c_x(\tau_0, \nu_0) & \Define & \int_{-\infty}^{\infty} p^*_{(\tau_0, \nu_0)}(t) \, x(t) \, dt
\end{eqnarray}
where the coefficient $c_x(\tau_0, \nu_0)$ corresponding to the basis signal $p_{(\tau_0, \nu_0)}(t)$
is the value of the ZAK representation of $x(t)$ at $\tau = \tau_0$ and $\nu = \nu_0$, i.e.
\begin{eqnarray}
\label{thmbasiseqn2}
c_x(\tau_0, \nu_0) & = & {\mathcal Z}_x(\tau_0, \nu_0).
\end{eqnarray}  
\end{theorem}
\begin{IEEEproof}
See Appendix \ref{prf_thmbasis}.
\end{IEEEproof}

\section{An Orthonormal Basis for Time and Bandwidth Limited Signals Which are Localized in DD Domain}
\label{secbw}
In this section we consider TD signals which are approximately time-limited to the interval $[0 \,,\, NT)$ and band-limited
to the interval $[0 \,,\, M \Delta f)$ where $M$ and $N$ are positive integers. We are specifically interested in those signals whose
ZAK representation is localized in the DD domain, since such signals can be used to modulate and demodulate information symbols in the DD domain
with little inter-symbol interference. These signals are also expected to be robust to channel induced Doppler shift, since from Result \ref{zlem52} we know that the effect of Doppler shift is to
only shift the signal along the Doppler domain.     

Although the TD signals $p_{(\tau_0, \nu_0)}(t), 0 \leq \tau_0 < T, 0 \leq \nu_0 < \Delta f$ in Lemma \ref{plem1st} are localized in the DD domain (see (\ref{lceqn1})),
they are neither time-limited nor band-limited. Therefore, we obtain another basis of approximately time and bandwidth limited signals, by approximately limiting
the signals $p_{(\tau_0, \nu_0)}(t), 0 \leq \tau_0 < T, 0 \leq \nu_0 < \Delta f$ in the time and frequency domain. For this, we firstly multiply each basis signal
$p_{(\tau_0, \nu_0)}(t)$ by a signal $q(t)$ which is approximately limited to the time-interval $[0 \,,\, NT)$, followed by convolution of the product signal $q(t) p_{(\tau_0, \nu_0)}(t)$
with another TD signal $s(t)$ which is approximately band-limited to the frequency domain interval $[ 0 \,,\, M \Delta f)$. These time- and band-limited
 signals are given by

{\vspace{-4mm}
\small
\begin{eqnarray}
\label{basiseqn1}
\psi^{(q,s)}_{(\tau_0, \nu_0)}(t) & \hspace{-3mm}  \Define &  \hspace{-3mm} \left( p_{(\tau_0, \nu_0)}(t) \, q(t) \right) \, \star  s(t),\, 0 \leq \tau_0 < T , \, 0 \leq \nu_0 < \Delta f, \nonumber \\
q(t) & \approx & 0 \,,\, t \notin [0 \,,\, NT), \nonumber \\
\left\vert {\mathcal F}_s(f) \right\vert & \hspace{-3mm}  = & \hspace{-3mm} \left\vert \int_{-\infty}^{\infty} \hspace{-2mm} s(t) e^{-j 2 \pi f t} \, dt \right\vert \, \approx \, 0 \,,\, f \notin [0 \,,\, M \Delta f),
\end{eqnarray}  
\normalsize} 
where $\star$ denotes the TD convolution operator. Although the signals $\psi^{(q,s)}_{(\tau_0, \nu_0)}(t), 0 \leq \tau_0 < T\,,\, 0 \leq \nu_0 < \Delta f,$ are approximately time and bandwidth limited, the degree of localization of these signals in the DD domain is not immediately obvious. Therefore, next we derive the ZAK representation of
these signals. 
Subsequently, in this paper we consider the ideal time-limited waveform
\begin{eqnarray}
\label{qteqn}
q(t) & \Define \begin{cases} 
1 &, \,\, 0 \leq t < NT \\
0 &, \,\, \mbox{\small{otherwise}}
\end{cases}
\end{eqnarray}
and the ideal band-limited waveform
\begin{eqnarray}
\label{steqn}
s(t) & \hspace{-3mm} = &  \hspace{-3mm} \int_{0}^{M \Delta f} \hspace{-3mm} e^{j 2 \pi f t} df  \,  =  \,  e^{j \pi M \Delta f t} \, M \Delta f  \, \mbox{\small{sinc}}(M \Delta f t), \nonumber \\
\mbox{\small{sinc}}(x) & \Define & \frac{\sin(\pi x)}{\pi x}.
\end{eqnarray}
Using (\ref{plemeqn2}), (\ref{qteqn}) and (\ref{steqn}) in (\ref{basiseqn1}), we get
\begin{eqnarray}
\label{basisqeqn}
\psi^{(q,s)}_{(\tau_0, \nu_0)}(t) & = &  \sqrt{T} \sum\limits_{n=0}^{N-1} e^{j 2 \pi \nu_0 n T} \, s(t - \tau_0 - nT)
\end{eqnarray}
where $s(t)$ is given by (\ref{steqn}).
From this expression, it is clear that due to bandwidth limitation, the train of impulses in $p_{(\tau_0, \nu_0)}(t)$ appears as train of $\mbox{\small{sinc}}(\cdot)$ pulses in $\psi^{(q,s)}_{(\tau_0, \nu_0)}(t)$, each $\mbox{\small{sinc}}(\cdot)$ pulse having width roughly twice the inverse bandwidth.
Further, due to time-limitation of $p_{(\tau_0, \nu_0)}(t)$,
this train of $\mbox{\small{sinc}}(\cdot)$ pulses is restricted to the interval $[ 0 \,,\, NT)$. In Fig.~\ref{fig3},
an illustration has been provided for $\frac{\sqrt{T}}{M} \left\vert \psi^{(q,s)}_{(\tau_0, \nu_0)}(t)  \right\vert$ when $M = 12, N=14$,
$\tau_0 = 0.6 T, \nu_0 = 0.2 \Delta f$.

The signal $p_{(\tau_0, \nu_0)}(t)$ is ideally localized in the DD domain.
However, we expect that time and bandwidth limitation of $p_{(\tau_0, \nu_0)}(t)$,
will affect its degree of localization in the DD domain.
In order to understand this, in the following theorem we derive the expression for the ZAK representation of $\psi^{(q,s)}_{(\tau_0, \nu_0)}(t)$.
\begin{theorem}
\label{thm21}
The ZAK representation of $\psi^{(q,s)}_{(\tau_0, \nu_0)}(t)$
is given by
\begin{eqnarray}
\label{eqn43}
{\mathcal Z}_{\psi, \tau_0, \nu_0}(\tau, \nu) & = & {\mathcal Z}_q\left( \tau_0 , \nu - \nu_0 \right) \, {\mathcal Z}_s\left( \tau - \tau_0, \nu \right)
\end{eqnarray}
where $ {\mathcal Z}_q(\tau,\nu)$ and ${\mathcal Z}_s(\tau,\nu)$ are the ZAK representations of $q(t)$ and $s(t)$.
These representations are given by
\begin{figure}[t]
\vspace{-0.3 cm}
\hspace{-0.3 in}
\centering
\includegraphics[width= 3.85 in, height= 2.9 in]{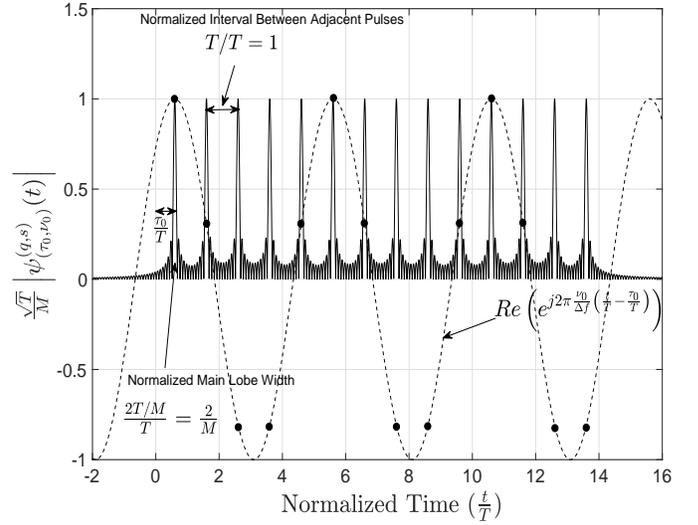}
\vspace{-0.2 cm}
\caption{ $\frac{\sqrt{T}}{M} \left\vert {\psi}_{(\tau_0, \nu_0)}^{(q,s)}(t) \right\vert$ vs. $t/T$ for $\frac{\tau_0}{T} = 0.6$, $\frac{\nu_0}{\Delta f} = 0.2$, $M=12, N=14$.} 
\vspace{-0.2cm}
\label{fig3}
\end{figure}

{\vspace{-4mm}
\small
\begin{eqnarray}
\label{expreqn1}
\hspace{-2mm} {\mathcal Z}_q(\tau, \nu) & \hspace{-3mm}  = &  \hspace{-3mm} \sqrt{T} e^{j 2 \pi \nu \left\lfloor \frac{\tau}{T} \right\rfloor T } e^{-j \pi \nu (N-1) T} \, \frac{\sin\left( \pi \nu N T \right)}{\sin\left( \pi \nu T \right)}, \nonumber \\
\hspace{-2mm} {\mathcal Z}_s(\tau, \nu) & \hspace{-3mm} = &  \hspace{-3mm} \frac{1}{\sqrt{T}} e^{j 2 \pi \nu \tau}  e^{-j 2 \pi \left\lfloor \frac{\nu}{\Delta f} \right\rfloor \Delta f \tau} \, e^{j \pi (M-1) \Delta f \tau} \frac{\sin( \pi M \Delta f \tau)}{\sin(\pi \Delta f \tau)}. \nonumber \\
\end{eqnarray}
\normalsize}
\end{theorem}
\begin{IEEEproof}
See Appendix \ref{prf_thm21}.
\end{IEEEproof}
Using the expressions for ${\mathcal Z}_q(\tau, \nu)$ and ${\mathcal Z}_s(\tau, \nu)$ from (\ref{expreqn1}) into the R.H.S. of (\ref{eqn43}), we get

{\vspace{-4mm}
\small
\begin{eqnarray}
\hspace{-4mm} \left\vert  {\mathcal Z}_{\psi, \tau_0, \nu_0}(\tau, \nu) \right\vert^2 & \hspace{-3mm}  = &  \hspace{-3mm} \frac{\sin^2\left( \pi (\nu - \nu_0) N T \right)}{\sin^2\left( \pi (\nu - \nu_0) T \right)} \, \frac{\sin^2( \pi M \Delta f (\tau - \tau_0))}{\sin^2(\pi \Delta f (\tau - \tau_0))}. \nonumber \\
\end{eqnarray}
\normalsize}
\begin{figure}[t]
\vspace{-0.5 cm}
\hspace{-0.1 in}
\centering
\includegraphics[width= 3.65 in, height= 2.9 in]{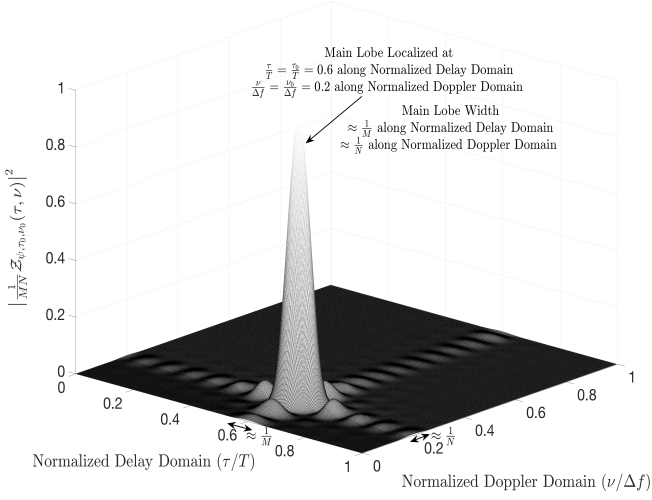}
\vspace{-0.2 cm}
\caption{$ \left\vert  \frac{1}{MN} {\mathcal Z}_{\psi, \tau_0, \nu_0}(\tau, \nu) \right\vert^2$ vs. $\left(\frac{\tau}{T}, \frac{\nu}{\Delta f} \right)$ for $\frac{\tau_0}{T} = 0.6$, $\frac{\nu_0}{\Delta f} = 0.2$, $M=12, N=14$.} 
\vspace{-0.2cm}
\label{fig4}
\end{figure}
From this expression it is clear that
the ZAK representation of $\psi^{(q,s)}_{(\tau_0, \nu_0)}(t)$ has most of its energy localized
around the point $(\tau_0, \nu_0)$ in the DD domain,
in an interval of width $1/(M \Delta f)$ and $\Delta f/N$ respectively along the delay and Doppler domains.
This is illustrated through Fig.~\ref{fig4} where we have plotted $\left\vert \frac{1}{MN}  {\mathcal Z}_{\psi, \tau_0, \nu_0}(\tau, \nu) \right\vert^2$
in the DD domain for $M = 12, N=14$, $\tau_0 = 0.6T, \nu_0 = 0.2 \Delta f$.
 This implies that two DD domain signals ${\mathcal Z}_{\psi, \tau_1, \nu_1}(\tau, \nu)$ and ${\mathcal Z}_{\psi, \tau_2, \nu_2}(\tau, \nu)$ (localized at $(\tau_1, \nu_1)$ and $(\tau_2, \nu_2)$ respectively), will not interfere significantly, if the points $(\tau_1, \nu_1)$ and $(\tau_2, \nu_2)$ are {\em separated} by roughly $1/(M \Delta f)$ along the delay domain and
by roughly $\Delta f/N$ along the Doppler domain (i.e., $\vert \tau_2 - \tau_1 \vert \approx 1/(M \Delta f)$ and $\vert \nu_2 - \nu_1 \vert \approx \Delta f/N$). 

Hence, for transmission of information, information symbols can linearly modulate
the DD domain signals ${\mathcal Z}_{\psi, \tau_0, \nu_0}(\tau, \nu)$. With sufficient separation in the DD domain, these DD domain signals will not interfere significantly
and therefore the information symbols can be recovered back from the modulated signal. Since $0 \leq \tau_0 < T, 0 \leq \nu_0 < \Delta f$ and the required separation along the delay and Doppler domains is at least $1/(M \Delta f)$ and $\Delta f/N$ respectively, we consider DD domain signals for $\tau_0 = l/(M \Delta f) = lT/M, l=0,1,\dots, M-1$ and
$\nu_0 = k \Delta f/N = k/(NT), k=0,1,\cdots, N-1$. From (\ref{steqn}) and (\ref{basisqeqn}), the $(k,l)$-th time-domain (TD) basis signal is then given by

{\vspace{-4mm}
\small
\begin{eqnarray}
\label{eqn27}
\alpha_{(k,l)}(t) & \Define & \frac{1}{\sqrt{MN}} \psi^{(q,s)}_{\left( \tau_0 = \frac{lT}{M}, \nu_0 = \frac{k}{NT} \right)}(t) \nonumber \\
& \hspace{-28mm}  \mya & \hspace{-16mm} \sqrt{\frac{T}{MN}} \sum\limits_{n=0}^{N-1}e^{j 2 \pi  \frac{ n  k}{N} } \,  \int_{0}^{M \Delta f} \hspace{-3mm} e^{j 2 \pi f \left( t - \frac{lT}{M} - nT \right)} df , \nonumber \\
& & k = 0,1,\cdots, N-1, \, l=0,1,\cdots, M-1,
\end{eqnarray}    
\normalsize}
where step (a) follows from (\ref{basisqeqn}) and the expression of $s(t)$ in (\ref{steqn}).
The factor $1/\sqrt{MN}$ ensures that these basis signals have unit energy.
Subsequently we denote this basis by $\boldsymbol {\alpha} = \left\{  \alpha_{(k,l)}(t) \right\}_{k=0,\cdots, N-1, l=0,1,\cdots,M-1}$.
The following theorem shows that $\boldsymbol {\alpha}$ is an $MN$-dimensional orthonormal basis.
\begin{theorem}
\label{lem18}
The basis $\boldsymbol {\alpha}$ is an $MN$-dimensional orthonormal basis, i.e.
\begin{eqnarray}
\label{lem18eqn}
\int_{-\infty}^{\infty} \hspace{-2mm} \alpha_{(k_1,l_1)}(t) \, \alpha^*_{(k_2,l_2)}(t) \, dt & = & \delta[k_1 - k_2] \, \delta[l_1 - l_2], \nonumber \\
& & \hspace{-45mm} k_1,k_2 = 0,1,\cdots, N-1 \,,\, l_1,l_2 = 0,1,\cdots, M-1.
\end{eqnarray}
\end{theorem}
\begin{IEEEproof}
See Appendix \ref{prf_lem18}.
\end{IEEEproof}
For a given $(M, N, T)$, we have therefore derived an $MN$-dimensional orthonormal basis $\boldsymbol {\alpha}$ of TD signals  which are localized in the DD domain, and which are approximately
time-limited to the interval $[0 \,,\, NT)$ and band-limited to the interval $[0 \,,\, M \Delta f)$. The dimensionality of the space of signals which are time-limited to $NT$ seconds and band-limited to $M \Delta f$ Hz but not necessarily localized in the DD domain, is known to be the time-bandwidth product $M \Delta f \times N T = MN$. Therefore, the additional constraint of localization
in the DD domain does not reduce the dimensionality of the space of time and bandwidth limited signals. 

\section{Delay-Doppler (DD) Domain Modulation}
\label{propmodulation}
From the discussion in the previous section, it is natural to consider DD domain modulation where the complex information symbols $x[k,l]$ linearly modulate the
basis signals $\alpha_{(k,l)}(t), k=0,1,\cdots, N-1, l=0,1,\cdots, M-1$. The TD transmit signal is therefore
given by

{\vspace{-4mm}
\small
\begin{eqnarray}
\label{eqn44}
x(t) & \hspace{-3mm} = & \hspace{-3mm} \sum\limits_{k=0}^{N-1}\sum\limits_{l=0}^{M-1} x[k,l] \, \alpha_{(k,l)}(t)  \, \mya  \,  \sqrt{\frac{T}{MN}} \sum\limits_{n=0}^{N-1}  x_n(t -nT) , \nonumber \\
x_n(t) & \hspace{-3mm}  \Define &  \hspace{-3mm}  \sum\limits_{k=0}^{N-1} \sum\limits_{l=0}^{M-1} x[k,l] \, e^{j 2 \pi \frac{n k}{N}} \int_{0}^{M \Delta f}  \hspace{-3mm} e^{j 2 \pi f \left( t - \frac{lT}{M}  \right)} \, df \nonumber \\
& \hspace{-3mm} = &  \hspace{-3mm} \int_{-\infty}^{\infty}  \hspace{-2mm} X_n(f)  \, e^{j 2 \pi f t} \, df  \,\,,\,\, n=0,1,\cdots, N-1, \nonumber \\
X_n(f) & \hspace{-3mm} \Define & \begin{cases}
   \sum\limits_{k=0}^{N-1} \sum\limits_{l=0}^{M-1}  x[k,l]  \, e^{j 2 \pi \left( \frac{n k}{N}  - \frac{f l T}{M} \right)}  , \, 0 \leq f < M \Delta f \\
   0 \, \hspace{38mm} , \, \mbox{\small{otherwise}}
\end{cases}, \nonumber \\
& &  n=0,1,\cdots, N-1
\end{eqnarray} 
\normalsize}
where in step (a) we have used the expression of $\alpha_{(k,l)}(t)$ from (\ref{eqn27}).
The expression of $x(t)$ in (\ref{eqn44}) suggests a two stage modulation method, where in the first stage the DD domain information symbols $x[k,l], k=0,1,\cdots, N-1, l=0,1,\cdots, M-1$ are transformed
to the signals $X_n(f), n=0,1,\cdots, N-1$. In the second stage, for each $n=0,1,\cdots, N-1,$ $X_n(f)$ is then transformed to the TD signal
$x_n(t)$ which is time-shifted by $nT$. These time-shifted signals are then added to get $x(t)$.

\section{Derivation of OTFS Modulation}
\label{deriveOTFS}
Using the DD domain modulation derived in Section \ref{propmodulation}, we derive the OTFS modulation equations in this section. 
To the best of our knowledge, all prior work on OTFS modulation/demodulation have only reported
the equations for transforming the DD domain information symbols to the TD transmit signal, and have not derived the modulation equations from first principles.
An important contribution of our current paper lies in the novel derivation of OTFS modulation equations from first principles.  

The following observation is useful in the derivation of OTFS modulation from (\ref{eqn44}).
In (\ref{eqn44}) we note that for any $(k,l), k=0,1,\cdots, N-1, l=0,1,\cdots, M-1$, the $(k,l)$-th term in the expression for $x_n(t)$ i.e., $x_{n,k,l}(t) \Define x[k,l] e^{j 2 \pi \frac{n k}{N}} \int_{0}^{M \Delta f}  \hspace{-2mm} e^{j 2 \pi f \left( t - \frac{lT}{M}  \right)} \, df =  M \Delta f \, x[k,l]  \, e^{j 2 \pi \frac{n k}{N}}  e^{j \pi  M \Delta f \left( t - \frac{lT}{M} \right)} \mbox{\small{sinc}}\left(M \Delta f \left( t - \frac{lT}{M}  \right) \right)$, has most of its energy in a time-interval of the type
$\left[ \frac{(l- \zeta)T}{M} \,,\, \frac{(l+\zeta)T}{M} \right]$, $\zeta \geq 1$. This is because $\left\vert  x_{n,k,l}(t) \right\vert^2 = (M \Delta f)^2 \vert x[k,l] \vert^2 \mbox{\small{sinc}}^2\left(M \Delta f \left( t - \frac{lT}{M}  \right) \right)$ and

{\vspace{-4mm}
\small
\begin{eqnarray}
\label{otfseqn87}
\hspace{-1mm} \frac{\int_{\frac{(l- \zeta)T}{M}}^{\frac{(l+ \zeta)T}{M}}  \left\vert x_{n,k,l}(t) \right\vert^2 dt}{\int_{-\infty}^{\infty}  \left\vert x_{n,k,l}(t) \right\vert^2 dt} & \hspace{-3mm} = &  \hspace{-3mm} 
 \frac{\int_{\frac{(l- \zeta)T}{M}}^{\frac{(l+ \zeta)T}{M}} (M \Delta f )^2 \mbox{\small{sinc}}^2\hspace{-1mm}\left(M \Delta f \left( t - \frac{lT}{M}  \right) \right) dt}{ {\int_{-\infty}^{\infty} (M \Delta f )^2 \mbox{\small{sinc}}^2\hspace{-1mm}\left(M \Delta f \left( t - \frac{lT}{M}  \right) \right) dt}}  \nonumber \\
 & \hspace{-3mm}  \mya & \hspace{-3mm}  \int_{-\zeta}^{\zeta}  \hspace{-2mm} \mbox{\small{sinc}}^2(t') dt' 
\end{eqnarray}
\normalsize}
where step (a) follows from the change of integration variable to $t' \Define M \Delta f \left(t - \frac{lT}{M} \right)$.
Assuming integer $\zeta  \geq 1$ and $M \geq 2 \zeta$, each term $x_{n,k,l}(t)$ corresponding to $l=
\zeta,\zeta + 1,\cdots, M- \zeta,$ has at least $\int_{-\zeta}^{\zeta} \mbox{\small{sinc}}^2(t') dt' $ fraction of its energy in the time-interval $[ 0 \,,\, T)$.
This follows from (\ref{otfseqn87}) and the fact that for each $l = \zeta,\zeta + 1,\cdots, M- \zeta,$ $\left[ \frac{(l - \zeta) T}{M} \,,\, \frac{(l+ \zeta)T}{M} \right) \subset [0 \,,\, T)$.
Next, for each $i=1,2,\cdots, \zeta -1$, the terms $x_{n,k,l}(t)$ corresponding to $l=i, M-i,$ have at least $\int_{-i}^{i} \mbox{\small{sinc}}^2(t') dt'$ fraction of their energy inside $[ 0 \,,\, T)$ (since for $l=i, M-i$, $\left[ \frac{(l-i)T}{M} \,,\, \frac{(l+i)T}{M}  \right) \, \subset \, [0 \,,\, T)$). However, the terms $x_{n,k,l}(t)$ corresponding to $l=0$
have at least half of their energy outside the interval $[ 0 \,,\, T)$. Since $x_n(t) = \sum\limits_{k=0}^{N-1} \sum\limits_{l=0}^{M-1} x_{n,k,l}(t)$, out of all the terms $x_{n,k,l}(t)$ corresponding to $l=0,1,\cdots, M-1$,
a fraction $(M- 2 \zeta + 1)/M$ correspond to $l=\zeta,\zeta + 1,\cdots, M- \zeta,$ and a fraction $2/M$ corresponds to $l=i, M-i$ for each $i=1,2,\cdots, \zeta -1$.
Therefore, for $M \geq 2 \zeta$, at least a fraction $\gamma(\zeta, M) \Define \left( \frac{M + 1 - 2\zeta}{M} \int_{-\zeta}^{\zeta}  \mbox{\small{sinc}}^2(t') dt' \, + \, \frac{2}{M} \sum\limits_{i=1}^{\zeta - 1} \int_{-i}^{i} \mbox{\small{sinc}}^2(t') dt' \right)$ of the total energy of $x_n(t)$ lies in the interval $[0 \,,\, T)$.
From the expression of this fraction $\gamma(\zeta, M)$ it is clear that for a fixed $\zeta \geq 1$, $\lim_{M \rightarrow \infty} \gamma(\zeta, M)  =    \int_{-\zeta}^{\zeta}  \mbox{\small{sinc}}^2(t') dt' $.
Since $\lim_{\zeta \rightarrow \infty} \int_{-\zeta}^{\zeta}  \mbox{\small{sinc}}^2(t') dt'  = \int_{-\infty}^{\infty}  \mbox{\small{sinc}}^2(t') dt' = 1$, it follows
that with a sufficiently large $M$, $x_n(t)$ has almost all of its energy in the time-interval $[0 \,,\, T)$.
 
The following theorem uses this observation to derive OTFS modulation equations from the DD domain modulation equations in (\ref{eqn44}).
Also, since $x_n(t), n=0,1,\cdots, N-1,$ are approximately time-limited to $[0 \,,\, T)$, it follows that the $n$-th term in step (a) of (\ref{eqn44}) i.e., $x_n(t - nT)$ is approximately time-limited to the interval $[ nT \,,\, (n+1)T)$. As $X_n(f)$ is the Fourier transform of $x_n(t)$, we subsequently refer to $X_n(f)$ as the $n$-th time-frequency (TF) signal.

\begin{theorem} (Derivation of OTFS modulation)
\label{thmotfsmod}
For sufficiently large $M$, the DD domain modulated signal $x(t)$ in (\ref{eqn44})
is given by

{\vspace{-4mm}
\small
\begin{eqnarray}
\label{otfsthmeqn}
\hspace{-8mm} x(t)  &  \hspace{-4mm} \myapproxa  &  \hspace{-4mm} \frac{1}{\sqrt{MN}} \sum\limits_{n=0}^{N-1} \sum\limits_{m=0}^{M-1} \hspace{-1mm} g(t - nT) \, X_{\mbox{\tiny{TF}}}[n,m] \, e^{j 2 \pi m \Delta f (t - nT)}, \nonumber \\
X_{\mbox{\tiny{TF}}}[n,m] & \hspace{-2mm}  \Define & \hspace{-2mm} \sum\limits_{k=0}^{N-1}\sum\limits_{l=0}^{M-1} x[k,l] \, e^{j 2 \pi \left( \frac{n k}{N} - \frac{m l}{M} \right)}, \nonumber \\
& &  \hspace{-1mm}  n=0,1,\cdots, N-1, m=0,1,\cdots, M-1, \nonumber \\
g(t) &  \hspace{-2mm} \Define &  \hspace{-2mm} \begin{cases}
 \frac{1}{\sqrt{T}}  \,,\, t \in [0 \,,\, T)  \\
 0  \,\,\,,\, \mbox{\small{otherwise}}
\end{cases}.
\end{eqnarray}
\normalsize}
The R.H.S. in step (a) is {\em exactly the equation for OTFS modulation} in \cite{HadaniOTFS1, HadaniOTFS2, HadaniOTFS3}, with a rectangular transmit pulse $g(t)$.
\end{theorem}
\begin{IEEEproof} 
Let us consider the product of $X_n(f)$ with the frequency domain pulse train $\Delta f  \hspace{-2mm} \sum\limits_{m=-\infty}^{\infty} \hspace{-2mm} \delta(f - m \Delta f)$, i.e.
\begin{eqnarray}
\label{otfsderiveqn7}
{\Tilde X_n}(f) & \Define & X_n(f) \, \left[ \Delta f  \hspace{-2mm} \sum\limits_{m=-\infty}^{\infty}  \hspace{-2mm} \delta(f - m \Delta f) \right].
\end{eqnarray}
Since the inverse Fourier transform of $\Delta f  \sum\limits_{m=-\infty}^{\infty} \delta(f - m \Delta f)$
is $ \sum\limits_{k=-\infty}^{\infty} \delta(t - k T)$, the inverse Fourier transform of ${\Tilde X_n}(f)$ (which we denote by
${\Tilde x_n}(t)$) is given by
\begin{eqnarray}
\label{otfsderiveqn5}
{\Tilde x_n}(t) & \hspace{-3mm}  = & \hspace{-3mm} x_n(t)  \star  \sum\limits_{k=-\infty}^{\infty} \hspace{-2mm} \delta(t - k T) = \sum\limits_{k=-\infty}^{\infty} \hspace{-2mm} x_n(t - kT) 
\end{eqnarray}
where $\star$ denotes TD convolution.
Note that ${\Tilde x_n}(t)$ is a periodic TD signal with period $T$.
As we have already observed, for a sufficiently large $M$, $x_n(t)$ is approximately time-limited to the interval $[0 \,,\, T)$.
Therefore, from (\ref{otfsderiveqn5}) it follows that
for $0 \leq t < T$ we have
\begin{eqnarray}
{\Tilde x_n}(t)  & \approx & x_n(t) \,\,,\,\, 0 \leq t < T.
\end{eqnarray}  
In other words
\begin{eqnarray}
\label{otfsderiveqn11}
x_n(t)  & \approx & \sqrt{T} \, g(t) \, {\Tilde x_n}(t), 
\end{eqnarray}
where $g(t)$ is the rectangular TD signal defined in (\ref{otfsthmeqn}).
Taking the inverse Fourier transform of ${\Tilde X_n}(f)$, from (\ref{otfsderiveqn7}) we get
\begin{eqnarray}
\label{otfsderiveqn8}
{\Tilde x_n}(t) & \hspace{-3mm}  = & \hspace{-3mm} \int_{-\infty}^{\infty} \hspace{-2mm} X_n(f) \, \Delta f  \hspace{-2mm} \sum\limits_{m=-\infty}^{\infty}  \hspace{-2mm} \delta(f - m \Delta f) \, e^{j 2 \pi ft} \, df \nonumber \\
& \hspace{-3mm}  \mya &   \hspace{-3mm} \int_{0}^{M \Delta f} \hspace{-2mm} X_n(f) \, \Delta f  \hspace{-2mm} \sum\limits_{m=-\infty}^{\infty}  \hspace{-2mm} \delta(f - m \Delta f) \, e^{j 2 \pi ft} \, df \nonumber \\
& = & \Delta f   \sum\limits_{m=0}^{M-1}   X_n(m \Delta f)  \, e^{j 2 \pi m \Delta f t}
\end{eqnarray}
where step (a) follows from the fact that $X_n(f)$ is limited to the frequency domain interval $[0 \,,\, M \Delta f)$ (see (\ref{eqn44})).
From the expression of $X_n(f)$ in (\ref{eqn44}) (in terms of the information symbols $x[k,l]$), for $m=0,1,\cdots, M-1$ we get
\begin{eqnarray}
\label{xnfmdfeqn}
X_n(m \Delta f) & =  & \sum\limits_{k=0}^{N-1} \sum\limits_{l=0}^{M-1}  x[k,l]  \, e^{j 2 \pi \left( \frac{n k}{N}  - \frac{(m \Delta f) l T}{M} \right)}  \nonumber \\
& \mya & \sum\limits_{k=0}^{N-1} \sum\limits_{l=0}^{M-1}  x[k,l]  \, e^{j 2 \pi \left( \frac{n k}{N}  - \frac{m l}{M} \right)} \nonumber \\
& \myb & X_{\mbox{\tiny{TF}}}[n,m],
\end{eqnarray}
where step (a) follows from the fact that $T \Delta f = 1$ and step (b) follows from the definition of
$X_{\mbox{\tiny{TF}}}[n,m]$ in (\ref{otfsthmeqn}). 
Using (\ref{xnfmdfeqn}) in (\ref{otfsderiveqn8}) we get
\begin{eqnarray}
\label{otfs11eqn2}
{\Tilde x_n}(t) =  \Delta f   \sum\limits_{m=0}^{M-1}   X_{\mbox{\tiny{TF}}}[n,m]  \, e^{j 2 \pi m \Delta f t}.
\end{eqnarray}
Using (\ref{otfs11eqn2}) in (\ref{otfsderiveqn11}) we get
\begin{eqnarray}
\label{apintgeqn}
x_n(t) &  \approx & \frac{1}{\sqrt{T}} \, g(t) \,  \sum\limits_{m=0}^{M-1}   X_{\mbox{\tiny{TF}}}[n,m] \, e^{j 2 \pi m \Delta f t}.
\end{eqnarray}
Using (\ref{apintgeqn}) in (\ref{eqn44}) we get
\begin{eqnarray}
x(t) &  \hspace{-3mm} = &  \hspace{-3mm} \sqrt{\frac{T}{MN}} \sum\limits_{n=0}^{N-1}  x_n(t -nT) \nonumber \\
&  & \hspace{-10mm} \myapproxa  \frac{1}{\sqrt{MN}} \sum\limits_{n=0}^{N-1} \sum\limits_{m=0}^{M-1} \hspace{-1mm} g(t - nT) \, X_{\mbox{\tiny{TF}}}[n,m] \, e^{j 2 \pi m \Delta f (t - nT)}, \nonumber \\
\end{eqnarray}
where we have used (\ref{apintgeqn}) in step (a).
This completes the proof.
\end{IEEEproof}

From (\ref{eqn44}) we know that the TD signal $x_n(t) = \int_{-\infty}^{\infty} X_n(f) e^{j 2 \pi f t} df$ is
the inverse Fourier transform of $X_n(f)$.
However, in practical systems it is difficult to exactly implement the integral in this inverse Fourier transform.
On the other hand, from (\ref{otfs11eqn2}) it is clear that ${\Tilde x}_n(t), 0 \leq t < T$
can be easily computed by an OFDM modulator in existing 4G/5G modems with a sub-carrier spacing of
$\Delta f$ and $M$ sub-carriers. For sufficiently large $M$, $x_n(t)$ is approximately time-limited to $[0 \,,\, T)$
and is approximately equal to ${\Tilde x}_n(t)$ in this interval (see (\ref{otfsderiveqn11})).
Therefore, for sufficiently large $M$, the DD domain modulation in (\ref{eqn44}) is same as OTFS modulation,
which can be practically implemented using the OFDM modulator in existing 4G/5G modems.

For sufficiently large $M$, we therefore expect the spectral efficiency (SE) performance of OTFS modulation to be same as that of the DD domain modulation in (\ref{eqn44}).
In the next section, we derive an expression for the SE achieved by the DD domain modulation in (\ref{eqn44}). 
Numerical simulations in Section \ref{secsim} reveal that, indeed the SE achieved by OTFS modulation is same as
the SE achieved by the DD domain modulation in (\ref{eqn44}). 
 
\section{Spectral Efficiency of DD Domain Modulation}
\label{Zakrecv}
In this section, we derive an expression for the spectral efficiency (SE) achieved by the DD domain modulation in (\ref{eqn44}).
For the derivation of the SE expression we consider a ZAK receiver  \cite{Saif2020}.  
With $x(t)$ as the time-domain (TD) transmit signal, the received TD signal is given by \cite{Bello}
\begin{eqnarray}
y(t)  & = &  \sum\limits_{i=1}^L h_i \, x(t - \tau_i) \, e^{j 2 \pi \nu_i (t - \tau_i)}  \, + \, n(t)
\end{eqnarray} 
where $h_i, \tau_i, \nu_i, i=1,2,\cdots,L$ are the channel gain, delay and Doppler shift of the $i$-th channel path between the transmitter and
the receiver. Also, $n(t)$ is the additive white Gaussian noise (AWGN) at the receiver.
Further, we consider $0 < \tau_i < T, \,  i=1,2,\cdots, L$.
Since, the transmit TD signal $x(t)$ has most of its energy in the time-interval $[0 \,,\, NT)$ and $0 < \tau_i < T, i=1,2,\cdots, L$,
the received signal $y(t)$ has most of its energy in the time-interval $[0 \,,\, (N+1)T)$.

In the ZAK receiver \cite{Saif2020}, the ZAK representation of the received TD signal i.e., ${\mathcal Z}_y(\tau, \nu)$ is sampled
at the discrete points $\left(\tau = \frac{l'T}{M}, \nu = \frac{k' \Delta f}{N}  \right)$ in the DD domain. This sampled received DD domain signal
is denoted by $Y[k',l'] \Define {\mathcal Z}_y\left(\tau = \frac{l'T}{M}, \nu = \frac{k' \Delta f}{N} \right), k'=0,1,\cdots, N-1, l'=0,1,\cdots, M-1$. 
The DD domain information
symbols $x[k,l], k=0,1,\cdots, N-1, l=0,1,\cdots, M-1$ are decoded from this sampled received DD domain signal.
The expression for $Y[k',l']$ in terms of the information symbols $x[k,l]$ is given by
the following theorem.
\begin{theorem}
\label{thm73}
The sampled received DD domain signal $Y[k',l']$
is given by (\ref{eqn67}) (see top of next page), where
${\mathcal Z}_n\left(\tau, \nu \right)$ is the ZAK representation of AWGN $n(t)$.
\end{theorem}
\begin{IEEEproof}
See Appendix \ref{prfthm73}.
\end{IEEEproof}
\begin{figure*}
\vspace{-8mm}
{\small
\begin{eqnarray}
\label{eqn67}
Y[k',l'] & = & \sqrt{MN} \sum\limits_{k=0}^{N-1} \sum\limits_{l=0}^{M-1}  x[k,l] \,  {\Tilde h}[k',l',k,l] \, + \, Z[k',l'], \,\, k'=0,1,\cdots, N-1, \, l' = ,1,\cdots, M-1, \nonumber \\
{\Tilde h}[k',l',k,l]  & \Define & \frac{1}{MN} \sum\limits_{i=1}^L h_i  {\Bigg \{} e^{j 2 \pi \frac{\nu_i}{\Delta f} \left(\frac{l'}{M} - \frac{\tau_i}{T}  \right)} e^{-j \pi (N- 1) \left( \frac{k'}{N} - \frac{k}{N} - \frac{\nu_i}{\Delta f} \right)}  e^{j 2 \pi \left( \frac{k'}{N} - \frac{\nu_i}{\Delta f} \right) \left( \frac{l'}{M} - \frac{\tau_i}{T} - \frac{l}{M}  \right)}  e^{-j 2 \pi \left\lfloor \frac{k'}{N} - \frac{\nu_i}{\Delta f}  \right\rfloor  \left(\frac{l'}{M} - \frac{\tau_i}{T} - \frac{l}{M}  \right)} \nonumber \\
& & \hspace{20mm} e^{j \pi (M-1) \left( \frac{l'}{M} - \frac{\tau_i}{T} - \frac{l}{M}   \right)}   \frac{\sin \left[ \pi N \left(\frac{k'}{N} - \frac{k}{N} - \frac{\nu_i}{\Delta f}  \right) \right] }{\sin \left[ \pi  \left(\frac{k'}{N} - \frac{k}{N} - \frac{\nu_i}{\Delta f}  \right) \right]} \, \frac{\sin \left[ \pi M \left(\frac{l'}{M} - \frac{l}{M} - \frac{\tau_i}{T}  \right) \right] }{\sin \left[ \pi  \left(\frac{l'}{M} - \frac{l}{M} - \frac{\tau_i}{T}  \right) \right]} {\Bigg \}}, \nonumber \\
Z[k',l'] & \Define & {\mathcal Z}_n\left(\tau = \frac{l'T}{M} , \nu = \frac{k' \Delta f}{N} \right).
\end{eqnarray}
\normalsize}
\begin{eqnarray*}
\hline
\end{eqnarray*}
\end{figure*}
The following theorem gives the expression for the SE achieved by the DD domain modulation in (\ref{eqn44}), with
a ZAK receiver.
\begin{theorem}
\label{thm8}
Let $\rho$ denote the ratio of the average transmit power to the received noise power (in the communication bandwidth $M \Delta f$). 
With i.i.d. complex Gaussian distributed information symbols, the SE achieved by the DD domain modulation in (\ref{eqn44}), with a ZAK receiver, is given by
\begin{eqnarray}
\label{Ceqn}
C & \Define & \frac{1}{MN} \, \log_2 \left\vert  {\bf I} \,  + \, \rho  \, {\Tilde {\bf H}}^H  {\Tilde {\bf K}}^{-1}  {\Tilde {\bf H}}  \right\vert,
\end{eqnarray}
where ${\bf I}$ denotes the $MN \times MN$ identity matrix and the elements of ${\Tilde {\bf H}}, {\Tilde {\bf K}} \in {\mathbb C}^{MN \times MN}$ are given by

{\vspace{-4mm}
\small
\begin{eqnarray}
{\Tilde  H}[k'M + l'+1, kM+l +1]  & \hspace{-3mm} \Define  \hspace{-3mm} & {\Tilde h}[k',l',k,l], \,\,\,\,\, (\mbox{\small{see (\ref{eqn67})}}) \nonumber \\
{\Tilde K}[{k'M + l' +1,kM + l +1}]  & \hspace{-3mm} =  & \hspace{-3mm} \begin{cases}
 \left(1 + \frac{1}{N} \right) , k' = k \,,\, l' = l \\
 \left( \frac{1}{N} \right) \hspace{6mm} , k' \ne k \,,\, l' = l \\
0  \hspace{11mm} \,, l' \ne l \\
\end{cases}, \nonumber \\
& & \hspace{-32mm} k',k = 0,1,\cdots, N-1 \,,\, l',l = 0,1,\cdots, M-1,
\end{eqnarray}
\normalsize}
where ${\Tilde h}[k',l',k,l]$ is given by (\ref{eqn67}). 
\end{theorem}
\begin{IEEEproof}
See Appendix \ref{prfthm8}.
\end{IEEEproof}

\section{Why is DD Domain Modulation Better Than OFDM ?}
\label{secbetter}
In this section, for the DD domain modulation in (\ref{eqn44}) and also for OFDM, we study the impact of channel induced Doppler shift on inter-symbol interference.
Let us consider a single path channel with delay $\tau'$, Doppler shift $\nu'$ and channel path gain $h'$.
Since we want to study inter-symbol interference, let us consider the channel to be noise-free.
From (\ref{eqn67}), the received DD domain samples in the ZAK receiver are then given by

{\vspace{-4mm}
\small
\begin{eqnarray}
\label{eqn91}
Y[k',l'] & = & \sqrt{MN} h'  \sum\limits_{k=0}^{N-1} \sum\limits_{l=0}^{M-1}  x[k,l] \,  R_{(k,l)}[k',l'], \nonumber \\
R_{(k,l)}[k',l']  &  \Define  &   \frac{{\Tilde h}[k',l',k,l]}{h'}, \nonumber \\
& &  \hspace{-6mm} k'=0,1,\cdots, N-1, \, l' = 0,1,\cdots, M-1.
\end{eqnarray}
\normalsize}
From (\ref{eqn91}) it is clear that the information symbol $x[k,l]$ is received in the $(k',l')$-th DD domain sample $Y[k',l']$
through the coefficient $R_{(k,l)}[k',l']$.    
From (\ref{eqn67})
it follows that

{\vspace{-4mm}
\small
\begin{eqnarray}
\label{eqn75}
\left\vert R_{(k,l)}[k',l'] \right\vert^2  & = & {\Bigg [}  \, \frac{\sin^2\left(  \pi \left(k' - k - \frac{\nu'}{\Delta f/N}    \right) \right)}{\sin^2\left( \frac{\pi}{N}  \left(k' - k - \frac{\nu'}{\Delta f/N}    \right) \right)}  \nonumber \\
& & \,\,\,\, \frac{\sin^2\left(  \pi \left(l' - l -  \tau' M \Delta f   \right) \right)}{\sin^2\left( \frac{\pi}{M}  \left(l' - l -  \tau' M \Delta f   \right) \right)}  {\Bigg ]}.
\end{eqnarray}
\normalsize}
From (\ref{eqn75}) it is clear that if the path delay $\tau'$ is an integer multiple of $1/(M \Delta f) = T/M$ and if the Doppler shift $\nu'$ is an integer multiple of $\Delta f/N = 1/(NT)$, then $\left\vert R_{(k,l)}[k',l'] \right\vert^2 = 0$
for all $k' \ne \left[ k + \frac{\nu'}{\Delta f/N} \right]_{_N} $ and $l ' \ne  \left[ l  +  \tau'M \Delta f  \right]_{_M} $, i.e.,
$x[k,l]$ is only received in $Y\left[k' = \left[ k + \nu'NT \right]_{_N}, l' = \left[ l  +  \tau'M \Delta f  \right]_{_M} \right]$
which is given by

{\vspace{-4mm}
\small
\begin{eqnarray}
\label{eqn83}
Y\left[k' = \left[ k + \nu'NT \right]_{_N}, l' = \left[ l  +  \tau'M \Delta f  \right]_{_M} \right]  & & \nonumber \\
& & \hspace{-70mm} = \sqrt{MN} h'   x[k,l]  R_{(k,l)} \hspace{-1mm} \left[k' = \left[ k + \nu'NT \right]_{_N} \hspace{-1.5mm} , l' = \left[ l  +  \tau'M \Delta f  \right]_{_M} \right], \nonumber \\
\end{eqnarray}
\normalsize}
i.e., there is no inter-symbol interference.
However, when the delay and Doppler shift are not integer multiples of $1/(M \Delta f)$ and $\Delta f/N$, $\left\vert R_{(k,l)}[k',l'] \right\vert^2$
is not zero for all $(k',l') \ne  \left( \left\lfloor \left[ k +  \nu'NT \right]_{_N}  \right\rfloor,  \left\lfloor \left[ l +  \tau'M \Delta f \right]_{_M}  \right\rfloor \right)$,
i.e., the energy of the DD domain basis signal carrying $x[k,l]$ leaks into other DD domain basis signals thereby creating inter-symbol interference.  
In the following we study the fraction of information symbols which can get significantly interfered by an information symbol.
\begin{figure}[t]
\vspace{-0.5 cm}
\hspace{-0.1 in}
\centering
\includegraphics[width= 3.55 in, height= 2.3 in]{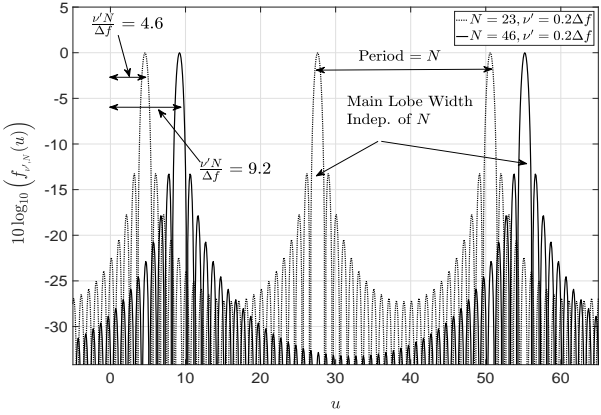}
\vspace{-0.3 cm}
\caption{$f_{_{\nu',N}}(u)$ vs. $u$, for $\frac{\nu'}{\Delta f} = 0.2$, $N=23, 46$.} 
\vspace{-0.3cm}
\label{fig5}
\end{figure}

Firstly, for a given $(k,l)$, from (\ref{eqn75}) we notice that $\left\vert R_{(k,l)}[k',l'] \right\vert^2, k'=0,1,\cdots, N-1 ,l'=0,1,\cdots,M-1$ is a product of two terms, one which depends on the difference
$(k' - k)$ along the Doppler domain and the other term which depends on the difference $(l' - l)$ along the delay domain.
To understand the spread/leakage of energy along the Doppler domain, we consider the 
function $f_{_{\nu',N}}(u)  \Define \frac{1}{N^2} \frac{\sin^2\left(  \pi \left(u - \frac{\nu'}{\Delta f/N}    \right) \right)}{\sin^2\left( \frac{\pi}{N}  \left(u - \frac{\nu'}{\Delta f/N}    \right) \right)}$.
Note that the term in $\left\vert R_{(k,l)}[k',l'] \right\vert^2$ which depends on $(k' - k)$ is given by $N^2 f_{_{\nu',N}}(u = k' - k)$.  
The function $f_{_{\nu',N}}(u)$ is periodic with period $N$ and has a peak at $u =  \frac{\nu'}{\Delta f/N}$ (see Fig.~\ref{fig5}).
From Fig.~\ref{fig5} and the expression for $f_{_{\nu',N}}(u)$ it is clear that the main lobe of $f_{_{\nu',N}}(u)$ is between $u =  \frac{\nu'}{\Delta f/N} - 1$ and $u =  \frac{\nu'}{\Delta f/N} +1$ (at both these values of
 $u$, $f_{_{\nu',N}}(u)$ is zero). Hence, the main lobe width of $f_{_{\nu',N}}(u)$ is two, which is independent of both $N$ and $\nu'$.
Therefore, along the Doppler domain, for a given $k$, $\left\vert R_{(k,l)}[k',l'] \right\vert^2$ is mostly localized
at $k' =  \left\lfloor \left[ k + s + \frac{\nu'}{\Delta f/N} \right]_{_N}  \right\rfloor, s=0,1$. Similarly, along the delay domain, for a given $l$, $\left\vert R_{(k,l)}[k',l'] \right\vert^2$
is mostly localized at $l'= \left\lfloor \left[ l  +  s + \tau'M \Delta f  \right]_{_M} \right\rfloor, s=0,1$.         
Hence, it follows that irrespective of $(\tau', \nu')$, in the DD domain, $\left\vert R_{(k,l)}[k',l'] \right\vert^2, k'=0,1,\cdots, N-1 ,l'=0,1,\cdots,M-1$
is mostly localized at $(k',l') \in {\mathcal A}_{(k,l)}$, where

{\vspace{-4mm}
\small
\begin{eqnarray}
{\mathcal A}_{(k,l)}  & \hspace{-3mm}   \Define  &   \hspace{-3mm} {\Bigg \{} (k' , l')\,  {\Bigg  \vert }  (k',l') \in {\mathcal S} \,\, \mbox{\small{and}}, k' =  \left\lfloor \left[ k + s_1 + \frac{\nu'}{\Delta f/N} \right]_{_N}  \right\rfloor \nonumber \\
& &  \hspace{-1mm}  \mbox{or} \,\,  l'= \left\lfloor \left[ l  +  s_2 + \tau'M \Delta f  \right]_{_M} \right\rfloor, \,\,  s_1, s_2 = 0,1 {\Bigg \}}, \nonumber \\
& &  \hspace{-12mm} {\mathcal S}  \Define  \{ (k',l') \, | \, k' = 0,1,\cdots,N-1, l'=0,1,\cdots, M-1 \}.
\end{eqnarray} 
\normalsize}
Therefore, the fraction
of interfered symbols (i.e., fraction of the remaining $MN -1 $ information symbols which receive significant interference from $x[k,l]$) is roughly $\frac{\left\vert  {\mathcal A}_{(k,l)}  \right\vert - 1}{MN - 1} = \frac{(2M + 2N -5)}{MN -1}$.
To be more precise, for a given $(k,l)$ and given $(\tau',\nu')$, let ${\mathcal B}_{(k,l)}$ denote the smallest cardinality set of $(k',l')$ pairs, such that
the sum of $\left\vert R_{(k,l)}[k',l']  \right\vert^2$ for all $(k',l') \in  {\mathcal B}_{(k,l)}$ is atleast $99 \%$ (i.e., $0.99$) of the total energy $\sum\limits_{k'=0}^{N-1} \sum\limits_{l'=0}^{M-1} \left\vert R_{(k,l)}[k',l']  \right\vert^2$, i.e.,

{\vspace{-2mm}
\small
\begin{eqnarray}
{\mathcal B}_{(k,l)} & \Define & \arg \hspace{-13mm} \min_{{\mathcal D}  \subseteq  {\mathcal S} \, {\Bigg \vert}  \,   \frac{\sum\limits_{(k',l') \in {\mathcal D} } \left\vert R_{(k,l)}[k',l']  \right\vert^2}{\sum\limits_{(k',l') \in {\mathcal S}} \left\vert R_{(k,l)}[k',l']  \right\vert^2 } \geq 0.99}  \hspace{-4mm} \left\vert {\mathcal D}  \right\vert.
\end{eqnarray}
\normalsize}
\begin{figure}[t]
\vspace{-0.2 cm}
\hspace{-0.1 in}
\centering
\includegraphics[width= 3.65 in, height= 2.5 in]{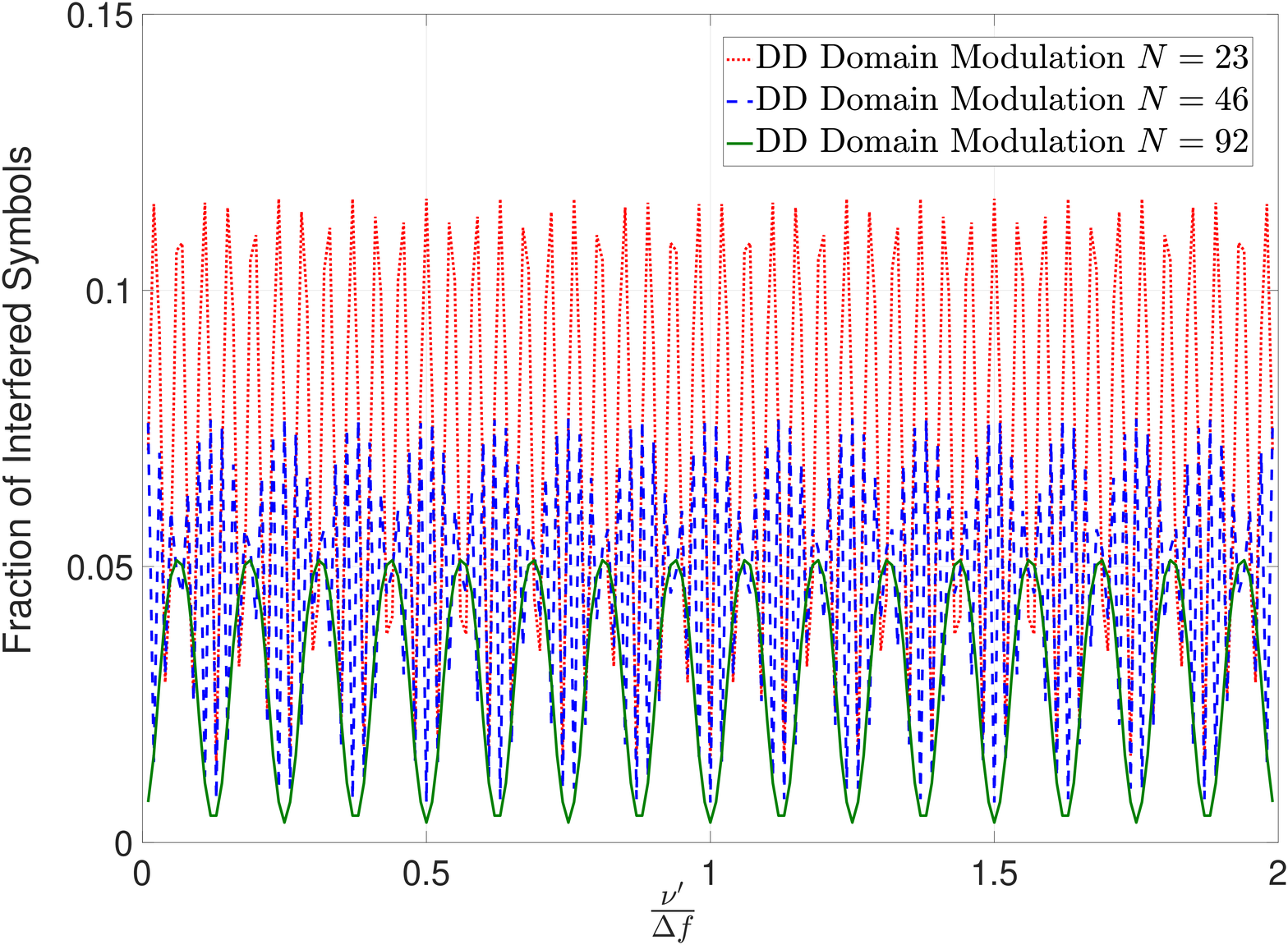}
\vspace{-0.4 cm}
\caption{Fraction of interfered DD domain information symbols vs. $\frac{\nu'}{\Delta f}$, for $N=23, 46, 92$, $M=45$.} 
\vspace{-0.2cm}
\label{fig6}
\end{figure}
In Fig.~\ref{fig6} we plot the fraction of interfered information symbols i.e. $\frac{\left(\left\vert  {\mathcal B}_{(k,l)} \right\vert - 1  \right)}{MN - 1}$
as a function of increasing $\nu'/\Delta f$, for $N=23,46, 92$ and $M=45, l = 23$. For each $\nu'/\Delta f$, the fraction of interfered symbols is averaged with respect to $\tau' M \Delta f$ which is uniformly distributed in the interval $[ 0 \,,\, 0.5]$. Further, $k=11,23, 46$, respectively for $N=23, 46, 92$. We observe that, for a given $N$ the fraction of interfered symbols is upper bounded for all values of $\nu'/\Delta f$. Further, this upper bound decreases with increasing $N$. For $N=23$, this upper bound is roughly $11.6 \%$, which decreases to $7.6 \%$ and $5.1 \%$ respectively for $N=46$ and $N=92$.
These values are close to our rough estimate of $\frac{(2M + 2N -5)}{MN -1}$, which is $12.67 \%$, $8.55 \%$ and $6.5 \%$ respectively for $N=23, 46, 92$.    

Next we compare the fraction of interfered information symbols in DD domain modulation with that 
in CP-OFDM (Cyclic-Prefix OFDM with sub-carrier spacing $\Delta f = 1/T$ and bandwidth $M \Delta f$).
In CP-OFDM it is well known that in the presence of channel induced Doppler shift, energy transmitted on a sub-carrier leaks into adjacent sub-carriers.
The fraction of interfered information symbols in CP-OFDM is presented in Appendix \ref{cpofdm}.  
\begin{figure}[t]
\vspace{-0.3 cm}
\hspace{-0.1 in}
\centering
\includegraphics[width= 3.65 in, height= 2.5 in]{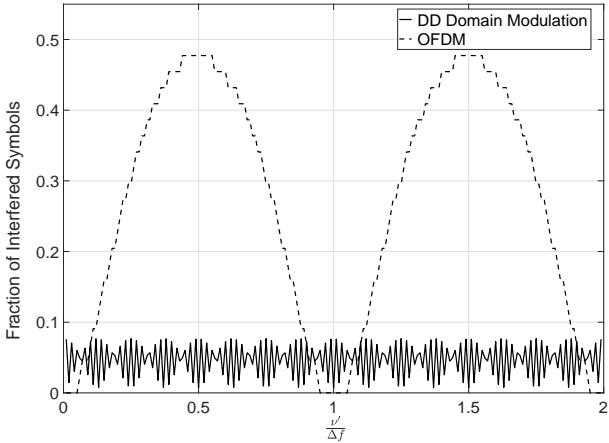}
\vspace{-0.4 cm}
\caption{Fraction of interfered information symbols vs. $\frac{\nu'}{\Delta f}$ for CP-OFDM and DD domain modulation.} 
\vspace{-0.2cm}
\label{fig7}
\end{figure}
In Fig.~\ref{fig7}, the fraction of interfered symbols for both CP-OFDM and DD domain modulation is plotted.
For CP-OFDM we have $M = 45$ and for DD domain modulation we have $M = 45, N = 46$, i.e., the bandwidth is $M \Delta f$ for
both CP-OFDM and DD domain modulation. The time duration of a CP-OFDM symbol is $1/\Delta f = T$, during which $M$ information symbols are transmitted, while
the time duration of a DD domain modulated signal is $NT$ during which $MN$ DD domain information symbols are transmitted.
In Fig.~\ref{fig7}, energy is transmitted on the $k=23$-rd sub-carrier in CP-OFDM, while in DD domain modulation, energy is transmitted on the $(k=23, l=23)$-th DD
domain basis signal. From the figure it is clear that the fraction of interfered symbols can be very large in CP-OFDM when compared to that in DD domain modulation.
For example, the maximum fraction of interfered symbols is $7.6 \%$ for DD domain modulation whereas it is about $48 \%$ for CP-OFDM.
Due to the significantly lower fraction of interfered symbols in DD domain modulation, it is practically feasible to perform joint DD domain equalization of all $MN$ information symbols
at the receiver.

Although a large $N$ helps in localizing the DD domain energy distribution of delay and Doppler shifted basis signals (which enables joint DD domain equalization), 
it also increases the duration of the modulated signal which is $NT$, i.e., $N$ times larger than the duration of an OFDM symbol.
In other words, {\em DD domain modulation achieves robustness to channel induced Doppler shift at the cost of
increased latency}.  

\section{Numerical Simulation}
\label{secsim}
In this section, we compare the spectral efficiency (SE) performance of the DD domain modulation in (\ref{eqn44}) and OTFS modulation.
We specifically consider communication of control and non-payload information between an Unmanned Aircraft System (UAS)
and a Ground Station (GS) \cite{ICTCPaper}. A widely accepted model for the en-route scenario
is the two-path model, with a direct and a reflected path. The gain for the direct path is $h_1 = \sqrt{K_f/(K_f+1)}$, while that for the reflected path is $h_2 \sim {\mathcal C}{\mathcal N}(0, 1/(K_f+1))$  \cite{Haas}. In \cite{Haas} it is mentioned that $K_f$ is typically $15$ dB. The delay between these two paths
is $33 \, \mu s$ (i.e., $\tau_1=0, \tau_2 = 33 \, \mu s$).
For an aircraft speed of $v$, the Doppler shift for the first path is $\nu_1 = v f_c/c$ and that for the second path is $\nu_2 = (v f_c/c) \cos\left(\pi - \theta \, {\mathcal U} \right)$,
where $f_c$ is the carrier frequency, $c = 3 \times 10^8$ m/s is the speed of light, ${\mathcal U}$ is a random variable uniformly distributed in the interval $[0 \,,\, 1]$ and $\theta = 3.5^{\circ}$ is the Doppler beamwidth \cite{Haas}. 
The channel bandwidth is $90$ KHz, time duration is $23$ ms, and $f_c = 5.06$ GHz \cite{ICTCPaper}. We therefore choose $M = 45, N = 46$ and $\Delta f = 2 $ KHz ($T = 1/\Delta f = 0.5$ ms, $M \Delta f = 45 \times 2$ KHz = $90$ KHz, $N T = 46 \times 0.5$ ms = $23$ ms).
\begin{figure}[t]
\vspace{-0.5 cm}
\hspace{-0.3 in}
\centering
\includegraphics[width= 3.65 in, height= 2.5 in]{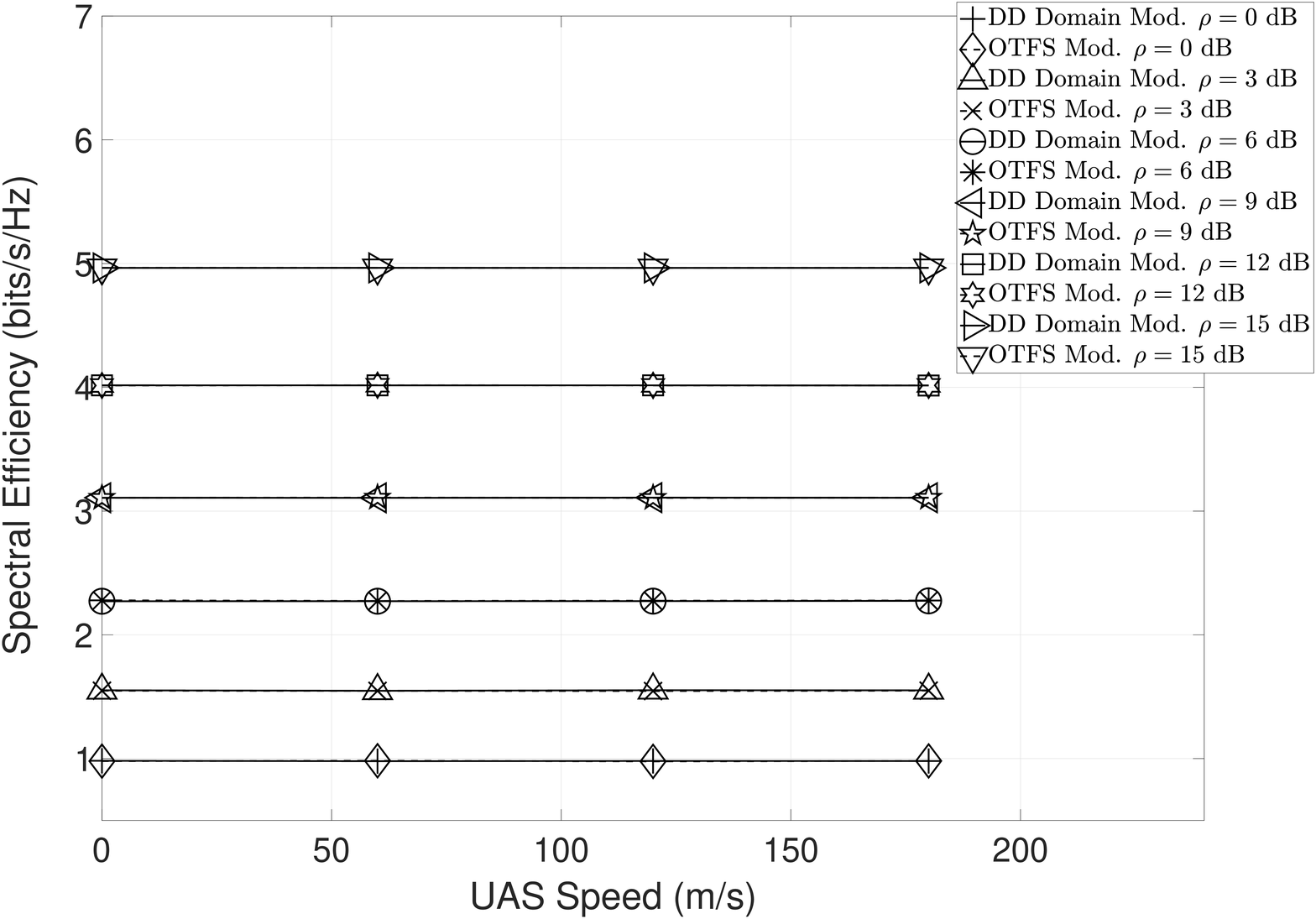}
\vspace{-0.2 cm}
\caption{Spectral Efficiency (SE) vs. aircraft speed (m/s) for DD Domain modulation in (\ref{eqn44}) and OTFS modulation.} 
\vspace{-0.2cm}
\label{fig8}
\end{figure}

In Fig.~\ref{fig8} we plot the average SE achieved by OTFS modulation and that achieved by the DD domain modulation in (\ref{eqn44}), as a function of increasing aircraft speed for different values of $\rho$.
The SE achieved by the DD domain modulation is given by ${\mathbb E}[C]$ (where $C$ is given by (\ref{Ceqn})) and that achieved by OTFS modulation is given by ${\mathbb E}[C_{\mbox{\tiny{Zak}}}]$ (where $C_{\mbox{\tiny{Zak}}}$ is given by equation $(24)$ in \cite{Saif2020}). 
Here, the expectation is w.r.t. the distribution of the channel path gains and the Doppler shift.
From Fig.~\ref{fig8}, it is clear that the SE performance of the DD domain modulation in (\ref{eqn44}) is same as that of OTFS modulation.
Further, for both these modulation schemes, for a given $\rho$, the SE performance remains constant with increasing aircraft speed (i.e., increasing Doppler shift).  

\section{Conclusion}
In this paper, using the ZAK representation of time-domain (TD) signals we have derived an orthonormal
basis of time and bandwidth limited signals which are also localized in the delay-Doppler (DD) domain.
We consider DD domain modulation based on this orthonormal basis and derive
Orthogonal Time Frequency Space (OTFS) modulation. To the best of our knowledge, this paper is the first
to rigorously derive OTFS modulation from first principles. We also show that with increasing time duration,
the basis signals are increasingly localized in the DD domain, irrespective of the amount of Doppler
shift. Increased localization reduces interference between information symbols
modulated on different basis signals, thereby enabling joint DD domain equalization of all information symbols.    
Therefore, DD domain modulation achieves robustness to Doppler shift at the cost of increased latency.
 
\appendices

\section{Proof of Result \ref{zlem1}}
\label{prf_zlem1}
From (\ref{zdef1}) it follows that
\begin{eqnarray}
{\mathcal Z}_x(\tau + T, \nu) & \hspace{-3mm} = \hspace{-3.5mm} & \sqrt{T}  \hspace{-1mm} \sum\limits_{k=-\infty}^{\infty} \hspace{-1mm} x(\tau + T + kT) \, e^{-j 2 \pi k \nu T } \nonumber \\
& \hspace{-3mm} = \hspace{-3.5mm} & e^{j 2 \pi \nu T} \sqrt{T} \hspace{-1mm} \sum\limits_{k=-\infty}^{\infty} \hspace{-2mm} x(\tau + (k+1)T ) e^{-j 2 \pi (k+1) \nu T } \nonumber \\
& \hspace{-3mm} = \hspace{-3.5mm} & e^{j 2 \pi \nu T}  {\mathcal Z}_x(\tau,\nu).
\end{eqnarray}
From (\ref{zdef1}) it also follows that
\begin{eqnarray}
{\mathcal Z}_x \left(\tau,\nu + \Delta f \right) & \hspace{-3mm}  = & \hspace{-3mm} \sqrt{T}  \sum\limits_{k=-\infty}^{\infty} \, x(\tau + kT) \, e^{-j 2 \pi k (\nu + \Delta f) T } \nonumber \\
& \hspace{-3mm}  \mya & \hspace{-3mm} \sqrt{T}  \sum\limits_{k=-\infty}^{\infty} \, x(\tau + kT) \, e^{-j 2 \pi k \nu T } \nonumber \\
& \hspace{-3mm}  = & \hspace{-3mm}  {\mathcal Z}_x \left(\tau,\nu \right) 
\end{eqnarray}
where step (a) follows from the fact that $T \Delta f = 1$.

\section{Proof of Result \ref{zlem3}}
\label{prf_zlem3}
From (\ref{zdef1}) it follows that
\begin{eqnarray}
\label{zeqn8}
\hspace{-4mm} \sqrt{T} \int\limits_{0}^{\Delta f} \hspace{-1mm} {\mathcal Z}_x(t, \nu) d\nu & \hspace{-3mm}  = & \hspace{-3mm} T  \hspace{-2mm} \sum\limits_{k=-\infty}^{\infty} \hspace{-1mm} \left[ x(t + kT)  \int\limits_{0}^{\Delta f} \hspace{-0.5mm} e^{-j 2 \pi \nu k T} d\nu \right].
\end{eqnarray}
Since $\Delta f = 1/T$ and $k \in {\mathbb Z}$ we have
\begin{eqnarray}
\label{zeqn9}
\int\limits_{0}^{\Delta f} \hspace{-0.5mm} e^{-j 2 \pi \nu k T} d\nu = 
\begin{cases}
0 &, k \ne 0   \\
\frac{1}{T} &, k = 0    \\
\end{cases}.
\end{eqnarray}
Using (\ref{zeqn9}) in (\ref{zeqn8}) we get (\ref{zeqn6}).
From (\ref{zdef1}) it also follows that

{\vspace{-4mm}
\small
\begin{eqnarray}
\label{zeqn10}
\frac{1}{\sqrt{T}} \hspace{-1mm} \int\limits_{0}^T \hspace{-1mm} {\mathcal Z}_x(\tau, f) e^{-j 2 \pi f \tau} d\tau & \hspace{-3.0mm}  = & \hspace{-4.5mm} \sum\limits_{k=-\infty}^{\infty} \int\limits_{0}^{T} \hspace{-1mm}  x(\tau + kT) e^{-j 2 \pi f (\tau + kT)} d\tau \nonumber \\
& & \hspace{-38mm} \mya   \sum\limits_{k=-\infty}^{\infty}  \hspace{-2mm} \int\limits_{kT}^{(k+1)T} \hspace{-2mm}  x(t) e^{-j 2 \pi f t} dt  =   \int\limits_{-\infty}^{\infty}  \hspace{-1mm} x(t) e^{-j 2 \pi f t} dt \, = \, {\mathcal F}_x(f)
\end{eqnarray}
\normalsize}
where step (a) follows from the substitution $t = \tau + kT$ in the integral in the previous step.

\section{Proof of Lemma \ref{plem1st}}
\label{prf_plem1st}
Using (\ref{zeqn6}) from Result \ref{zlem3}, the TD signal $p_{(\tau_0, \nu_0)}(t)$
having ZAK representation ${\mathcal Z}_{(p, \tau_0, \nu_0)}(\tau, \nu)$ in (\ref{lceqn1}) is given by 

{\vspace{-4mm}
\small
\begin{eqnarray}
p_{(\tau_0,\nu_0)}(t) & \hspace{-3mm} = & \hspace{-3mm} \sqrt{T} \int_{0}^{\Delta f} \hspace{-1mm} {\mathcal Z}_{(p, \tau_0, \nu_0)}(t, \nu) \, d\nu \nonumber \\
& \hspace{-32mm} \mya &   \hspace{-15.5mm} \sqrt{T} \hspace{-2mm}  \sum\limits_{n=-\infty}^{\infty} \hspace{-3mm} e^{j 2 \pi \nu_0 n T} \delta(t - \tau_0 - nT) \int_{0}^{\Delta f} \hspace{-3mm} \sum\limits_{m=-\infty}^{\infty} \hspace{-3mm} \delta (\nu - \nu_0 - m \Delta f)  d\nu \nonumber \\
& \hspace{-32mm} = &   \hspace{-15.5mm} \sqrt{T} \hspace{-2mm}  \sum\limits_{n=-\infty}^{\infty} \hspace{-3mm} e^{j 2 \pi \nu_0 n T} \delta(t - \tau_0 - nT) 
\end{eqnarray}
\normalsize}
where step (a) follows from the expression for ${\mathcal Z}_{(p, \tau_0, \nu_0)}(\tau, \nu)$ in (\ref{lceqn1}).

\section{Proof of Theorem \ref{thmbasis}}
\label{prf_thmbasis}
Firstly, we see that

{\vspace{-4mm}
\small
\begin{eqnarray}
\label{prfceqn}
c_x(\tau_0, \nu_0) & \hspace{-3mm} = &  \hspace{-3mm} \int_{-\infty}^{\infty} p^*_{(\tau_0, \nu_0)}(t) \, x(t) \, dt \nonumber \\
& \hspace{-3mm} \mya & \hspace{-3mm} \sqrt{T} \sum\limits_{n=-\infty}^{\infty} e^{-j 2 \pi \nu_0 n T} \int_{-\infty}^{\infty} \delta(t - \tau_0 - nT ) \, x(t) \, dt \nonumber \\
& \hspace{-3mm} = &  \hspace{-3mm} \sqrt{T} \sum\limits_{n=-\infty}^{\infty} e^{-j 2 \pi \nu_0 n T} \, x(\tau_0 + nT) \, \myb \, {\mathcal Z}_x(\tau_0, \nu_0)
\end{eqnarray}
\normalsize}
where step (a) follows from the expression of $p_{(\tau_0, \nu_0)}(t)$ in (\ref{plemeqn2}) and step (b) follows from (\ref{zdef1}).
Next

{\vspace{-4mm}
\small
\begin{eqnarray}
\int_{0}^T  \hspace{-2mm} \int_{0}^{\Delta f} \hspace{-4.5mm} c_x(\tau_0, \nu_0) p_{(\tau_0, \nu_0)}(t) d\tau_0 \, d\nu_0  & \hspace{-3mm}  \mya  &  \hspace{-4mm} \int_{0}^T  \hspace{-2mm} \int_{0}^{\Delta f} \hspace{-4.5mm} {\mathcal Z}_x(\tau_0, \nu_0) p_{(\tau_0, \nu_0)}(t) d\tau_0 d\nu_0 \nonumber \\
& \hspace{-94mm} \myb & \hspace{-48mm}   \hspace{-2mm}  \sum\limits_{n_1 = -\infty}^{\infty}  \sum\limits_{n_2 = -\infty}^{\infty}  {\Bigg \{}  \left[ \int_{0}^T \hspace{-1.5mm} x(\tau_0 + n_1 T) \, \delta(t - \tau_0 - n_2 T) \, d\tau_0  \right] \nonumber \\
& &  \hspace{-7mm} \underbrace{\left[ T \int_{0}^{\Delta f}  \hspace{-2mm} e^{j 2 \pi \nu_0 (n_2 - n_1) T} d\nu_0 \right]}_{= 1 \, \mbox{\tiny{if}} \, n_2 = n_1, 0 \, \mbox{\tiny{otherwise}} } {\Bigg \}} \nonumber \\
& \hspace{-94mm} = & \hspace{-48mm}  \sum\limits_{n_1 = -\infty}^{\infty}  \int_{0}^T \hspace{-1.5mm} x(\tau_0 + n_1 T) \, \overbrace{\delta(t - \tau_0 - n_1 T)}^{\mbox{\tiny{Dirac-delta at $\tau_0 = t - n_1 T$}}}  \, d\tau_0 \nonumber \\
& \hspace{-94mm} \myc & \hspace{-48mm}   \int_{0}^T \hspace{-1.5mm} x\left(\tau_0 + \left\lfloor \frac{t}{T} \right\rfloor T \right) \, \overbrace{\delta\left(t - \tau_0 -  \left\lfloor \frac{t}{T} \right\rfloor T \right)}^{\mbox{\tiny{Dirac-delta at $\tau_0 = t - \left\lfloor \frac{t}{T}  \right\rfloor T$}}}  \, d\tau_0  \, = \, x(t) \nonumber \\
\end{eqnarray}
\normalsize}
where step (a) follows from (\ref{prfceqn}). Step (b) follows from substituting the expressions for ${\mathcal Z}_x(\tau_0, \nu_0)$ from (\ref{zdef1}) and the
expression for $p_{(\tau_0, \nu_0)}(t)$ from (\ref{plemeqn2}) into the R.H.S. in step (a). Step (c) follows from the fact that the only non-zero term in the summation in the previous step is
for $n_1 = \left\lfloor \frac{t}{T} \right\rfloor$.

\section{ZAK Representation of Convolution of TD Signals}
\label{prf_zlem5}
This result states that the ZAK representation of the convolution of two time-domain (TD) signals is equivalent to convolution of their
ZAK representations in the delay domain.
\begin{result} \mbox{[see $(4.2)$ in \cite{Janssen}]}
\label{zlem5}
Let ${\mathcal Z}_a(\tau,\nu)$ and ${\mathcal Z}_b(\tau,\nu)$ be the ZAK representation of $a(t)$ and $b(t)$ respectively.
Consider their TD convolution $c(t) = \int\limits_{-\infty}^{\infty} a(t') \, b(t - t') \, dt'$. The ZAK representation of $c(t)$ is given by

{\vspace{-4mm}
\small
\begin{eqnarray}
\label{zeqn12}
{\mathcal Z}_c(\tau, \nu) & = & \frac{1}{\sqrt{T}} \int\limits_{0}^T {\mathcal Z}_a(\tau - \tau',\nu) \, {\mathcal Z}_b(\tau' , \nu) \, d\tau' \nonumber \\
& = & \frac{1}{\sqrt{T}} \int\limits_{0}^T {\mathcal Z}_a(\tau',\nu) \, {\mathcal Z}_b(\tau - \tau' , \nu) \, d\tau'.
\end{eqnarray}
\normalsize}
\end{result}
\begin{IEEEproof}
Starting with the R.H.S. of (\ref{zeqn12}) we have

{\vspace{-4mm}
\small
\begin{eqnarray}
\frac{1}{\sqrt{T}} \int\limits_{0}^T {\mathcal Z}_a(\tau - \tau',\nu) {\mathcal Z}_b(\tau' , \nu) d\tau'  &  \mya & \nonumber \\
& & \hspace{-55mm} \sqrt{T}  \hspace{-2mm} \sum\limits_{k_1=-\infty}^{\infty}  \sum\limits_{k_2=-\infty}^{\infty}  \hspace{-2mm} e^{- j 2 \pi (k_1 + k_2) \nu T} \hspace{-1mm} \int\limits_{0}^T a(\tau - \tau' + k_1 T) b(\tau' + k_2 T) d\tau' \nonumber \\
& \hspace{-98mm}  \myb &  \hspace{-52.5mm}  \sqrt{T}  \hspace{-2mm} \sum\limits_{k_1=-\infty}^{\infty}  \sum\limits_{k_2=-\infty}^{\infty}  \hspace{-3mm} e^{- j 2 \pi (k_1 + k_2) \nu T} \hspace{-4.5mm} \int\limits_{k_2 T}^{(k_2 + 1)T} \hspace{-4.5mm} a(\tau + (k_1 +k_2) T - \tau'' ) b(\tau'') d\tau'' \nonumber \\
& \hspace{-98mm}  \myc &  \hspace{-52.5mm}  \sqrt{T}  \hspace{-2mm} \sum\limits_{k_3=-\infty}^{\infty}   \hspace{-3mm} e^{- j 2 \pi k_3 \nu T} \left[ \sum\limits_{k_2=-\infty}^{\infty}  \hspace{-2.5mm} \int\limits_{k_2 T}^{(k_2 + 1)T} \hspace{-4.5mm} a(\tau + k_3 T - \tau'' ) b(\tau'') d\tau'' \right] \nonumber \\
& \hspace{-98mm} =   &  \hspace{-52.5mm}  \sqrt{T}  \hspace{-2mm} \sum\limits_{k_3=-\infty}^{\infty}   \hspace{-3mm} e^{- j 2 \pi k_3 \nu T} \left[  \int\limits_{-\infty}^{\infty} \hspace{-1.5mm} a(\tau + k_3 T - \tau'' ) b(\tau'') d\tau'' \right] \nonumber \\
& \hspace{-98mm} \myd   &  \hspace{-52.5mm}  \sqrt{T}  \hspace{-2mm} \sum\limits_{k_3=-\infty}^{\infty}   \hspace{-3mm} e^{- j 2 \pi k_3 \nu T} c(\tau + k_3 T ) \, = \, {\mathcal Z}_c(\tau, \nu)
\end{eqnarray}
\normalsize}
where step (a) follows from (\ref{zdef1}). In step (b) we have changed the integration variable to $\tau'' = \tau' + k_2 T$.
Step (c) follows from replacing the summation variable $k_1$ by $k_3 = k_1 + k_2$ since in step (b), $k_1$ always appears as $(k_1 + k_2)$.
Step (d) follows from the fact that $c(t)$ is the convolution of $a(t)$ and $b(t)$, i.e., $c(t) = \int\limits_{-\infty}^{\infty} a(t - t'') b(t'') dt''$.
The last step follows from (\ref{zdef1}). The second equality in the R.H.S. of (\ref{zeqn12}) can be proved similarly.
\end{IEEEproof}

\section{ZAK Representation of Product of TD Signals}
\label{prf_zlem6}
This result states that the ZAK representation of the product of two time-domain (TD) signals is equivalent to convolution of their
ZAK representations in the Doppler domain.
\begin{result} \mbox{[see $(4.4)$ in \cite{Janssen}]}
\label{zlem6}
Let ${\mathcal Z}_a(\tau,\nu)$ and ${\mathcal Z}_b(\tau,\nu)$ be the ZAK representation of TD signals $a(t)$ and $b(t)$ respectively.
Consider their product $c(t) = a(t) b(t)$. The ZAK representation of $c(t)$ is given by

{\vspace{-4mm}
\small
\begin{eqnarray}
\label{zeqn13}
{\mathcal Z}_c(\tau, \nu) & = &  \sqrt{T} \int\limits_{0}^{\Delta f} {\mathcal Z}_a(\tau,\nu - \nu') {\mathcal Z}_b(\tau , \nu') d\nu' \nonumber \\
& = &  \sqrt{T} \int\limits_{0}^{\Delta f} {\mathcal Z}_a(\tau,\nu') {\mathcal Z}_b(\tau, \nu - \nu') d\nu'.
\end{eqnarray}
\normalsize}
\end{result}
\begin{IEEEproof}
Starting with the R.H.S. in (\ref{zeqn13}) we get

{\vspace{-4mm}
\small
\begin{eqnarray}
\sqrt{T} \int\limits_{0}^{\Delta f} {\mathcal Z}_a(\tau,\nu - \nu') {\mathcal Z}_b(\tau , \nu') d\nu'  &  &   \nonumber \\
& & \hspace{-44mm} \mya \, \sqrt{T}   \hspace{-2mm} \sum\limits_{k_1=-\infty}^{\infty}  \sum\limits_{k_2=-\infty}^{\infty}  \hspace{-1mm} {\Bigg [} a(\tau + k_1 T) b(\tau + k_2 T) e^{-j 2 \pi \nu k_1 T}  \nonumber \\
& &  \hspace{-6mm} \underbrace{\frac{1}{\Delta f} \int_{0}^{\Delta f} e^{j 2 \pi \nu' (k_1 - k_2) T} d\nu' }_{= 1 \, \mbox{\tiny{if}} \, k_1 = k_2, \, 0 \, \mbox{\tiny{otherwise}}}  {\Bigg ]} \nonumber \\
&  \hspace{-88mm} \myb & \hspace{-46mm} \sqrt{T}   \hspace{-2mm} \sum\limits_{k_1=-\infty}^{\infty}  \hspace{-1mm} \underbrace{a(\tau + k_1 T) b(\tau + k_1T)}_{= c(\tau + k_1T)} e^{-j 2 \pi \nu k_1 T} \, \myc \, {\mathcal Z}_c(\tau,\nu)
\end{eqnarray}
\normalsize}
where step (a) follows from (\ref{zdef1}).
Step (b) follows from the fact that in the double summation in the previous step, only those terms are non-zero for which $k_2 = k_1$. Step (c) follows
from (\ref{zdef1}) and the fact that $c(t) = a(t) b(t)$.
\end{IEEEproof}

\section{Proof of Theorem \ref{thm21}}
\label{prf_thm21}
Using Result \ref{zlem6} in Appendix \ref{prf_zlem6}, for $0 \leq \tau < T$, the ZAK representation of
$c(t) = p_{(\tau_0, \nu_0)}(t) \, q(t)$ is given by

{\vspace{-4mm}
\small
\begin{eqnarray}
\label{zceqn23}
{\mathcal Z}_c(\tau, \nu) & = & \sqrt{T} \int_{0}^{\Delta f} {\mathcal Z}_{(p,\tau_0, \nu_0)}(\tau, \nu') \, {\mathcal Z}_q(\tau, \nu - \nu') \, d\nu' \nonumber \\
& \mya &  \sqrt{T} \int_{0}^{\Delta f}   \delta(\tau - \tau_0) \delta(\nu' - \nu_0) \, {\mathcal Z}_q(\tau, \nu - \nu') \,  d\nu' \nonumber \\
& = & \sqrt{T}  \, \delta(\tau - \tau_0) \, {\mathcal Z}_q(\tau, \nu - \nu_0)
\end{eqnarray}
\normalsize}
where step (a) follows from the fact that ${\mathcal Z}_{(p,\tau_0, \nu_0)}(\tau, \nu)$ is the ZAK representation of $p_{(\tau_0, \nu_0)}(t)$ (see Lemma \ref{plem1st}), and also that, for $0 \leq \tau < T$ and $0 \leq \nu' < \Delta f$, ${\mathcal Z}_{(p, \tau_0, \nu_0)}(\tau, \nu') = \delta(\tau - \tau_0) \delta(\nu' - \nu_0)$
(see (\ref{lceqn1})).
Here ${\mathcal Z}_q(\tau, \nu)$ is the ZAK representation of $q(t)$, and is given by

{\vspace{-4mm}
\small
\begin{eqnarray}
\label{zqeqn1}
{\mathcal Z}_q(\tau, \nu) & \hspace{-3mm}  = & \hspace{-3mm}  \sqrt{T} \hspace{-1mm} \sum\limits_{n=-\infty}^{\infty} \hspace{-2mm} q(\tau + nT) \, e^{-j 2 \pi \nu n T}  \, \mya  \, \sqrt{T}  \hspace{-1mm} \sum\limits_{n=- \left\lfloor \frac{\tau}{T} \right\rfloor }^{N-1 - \left\lfloor \frac{\tau}{T} \right\rfloor} \hspace{-2.5mm} e^{-j 2 \pi \nu n T} \nonumber \\
& \hspace{-3mm}  = &  \hspace{-3mm}  \sqrt{T} e^{j 2 \pi \nu \left\lfloor \frac{\tau}{T} \right\rfloor T } e^{-j \pi \nu (N-1) T} \, \frac{\sin\left( \pi \nu N T \right)}{\sin\left( \pi \nu T \right)}
\end{eqnarray}
\normalsize}
where step (a) follows from the fact that $q(t) = 1$ for $0 \leq t < NT$ and is zero otherwise (see (\ref{qteqn})).
From (\ref{basiseqn1}) it follows that
$\psi^{(q,s)}_{(\tau_0, \nu_0)}(t) = \left( p_{(\tau_0, \nu_0)}(t) \, q(t) \right) \star s(t) = c(t) \star s(t)$ and therefore using Result \ref{zlem5} from Appendix \ref{prf_zlem5}
it follows that

{\vspace{-4mm}
\small
\begin{eqnarray}
{\mathcal Z}_{\psi, \tau_0, \nu_0}(\tau, \nu) & \hspace{-3mm}  = &  \hspace{-3mm} \frac{1}{\sqrt{T}}  \int_{0}^T  {\mathcal Z}_c(\tau', \nu) \, {\mathcal Z}_s(\tau - \tau' , \nu) \, d\tau' \nonumber \\
&  \hspace{-14mm} \mya &  \hspace{-9mm} \int_{0}^{T} \delta(\tau' - \tau_0) \, {\mathcal Z}_q(\tau', \nu - \nu_0) \, {\mathcal Z}_s(\tau - \tau' , \nu) \, d\tau' \nonumber \\
& = & {\mathcal Z}_q(\tau_0, \nu - \nu_0) \, {\mathcal Z}_s(\tau - \tau_0 , \nu)
\end{eqnarray}
\normalsize}
where step (a) follows from the expression of ${\mathcal Z}_c(\tau', \nu)$ in (\ref{zceqn23}) for $0 \leq \tau' < T$.
Here ${\mathcal Z}_s(\tau,\nu)$ is the ZAK representation of $s(t)$ which is given by

{\vspace{-4mm}
\small
\begin{eqnarray}
{\mathcal Z}_s(\tau, \nu) & = & \sqrt{T} \sum\limits_{n=-\infty}^{\infty} s(\tau + nT) \, e^{-j 2 \pi \nu n T} \nonumber \\
& \mya & \sqrt{T} \int_{0}^{M \Delta f} \hspace{-2mm} e^{j 2 \pi f \tau } \underbrace{\left[ \sum\limits_{n=-\infty}^{\infty}  \hspace{-2mm}  e^{j 2 \pi (f - \nu) n T}  \right]}_{= \frac{1}{T} \hspace{-2mm} \sum\limits_{m = - \infty}^{\infty} \delta(f - \nu - m \Delta f)  } \, df \nonumber \\
& \myb & \frac{1}{\sqrt{T}}  \sum\limits_{m=-\infty}^{\infty} \int_{0}^{M \Delta f} \hspace{-4mm} \underbrace{\delta(f - \nu - m \Delta f)}_{\mbox{\small{Dirac-delta at}} \,  f = \nu + m \Delta f} \hspace{-3mm}  e^{j 2 \pi f \tau }  df. \nonumber \\
& \myc & \frac{1}{\sqrt{T}} e^{j 2 \pi \nu \tau} \sum\limits_{m= -\left\lfloor \frac{\nu}{\Delta f} \right\rfloor }^{(M-1) - \left\lfloor \frac{\nu}{\Delta f} \right\rfloor} e^{j 2 \pi m \Delta f \tau} \nonumber \\
&  & \hspace{-20mm} = \frac{1}{\sqrt{T}} e^{j 2 \pi \nu \tau}  e^{-j 2 \pi \left\lfloor \frac{\nu}{\Delta f} \right\rfloor \Delta f \tau} \, e^{j \pi (M-1) \Delta f \tau} \, \frac{\sin( \pi M \Delta f \tau)}{\sin(\pi \Delta f \tau)}
\end{eqnarray}
\normalsize}
where step (a) follows from the fact that $s(t) = \int_{0}^{M \Delta f} e^{j 2 \pi f t} df$ (see (\ref{steqn})).
Step (b) follows from the standard equation $\sum\limits_{n=-\infty}^{\infty}  \hspace{-2mm}  e^{j 2 \pi f n T} = \frac{1}{T} \hspace{-2mm} \sum\limits_{m = - \infty}^{\infty}  \hspace{-3mm} \delta\left(f - \frac{m}{T} \right)$ and the fact that $T = 1/\Delta f$. Step (c) follows from the fact that $0 \leq \nu + m \Delta f < M \Delta f$ for $- \left\lfloor \frac{\nu}{\Delta f} \right\rfloor \leq m \leq (M-1) - \left\lfloor \frac{\nu}{\Delta f} \right\rfloor$.
Indeed, using the R.H.S. above as the expression for ${\mathcal Z}_s(\tau, \nu)$, from (\ref{zeqn7}) it can be checked that the Fourier transform of $s(t)$ i.e., ${\mathcal F}_s(f) =  \frac{1}{\sqrt{T}} \int\limits_{0}^T {\mathcal Z}_s(\tau, f) e^{-j 2 \pi f \tau} \, d\tau$
is one when $f \in [0 \,,\, M \Delta f)$ and is zero otherwise.

\section{Proof of Theorem \ref{lem18}}
\label{prf_lem18}
From the expression of $\alpha_{(k,l)}(t)$ in (\ref{eqn27}) we have

{\vspace{-4mm}
\small
\begin{eqnarray}
 \int_{-\infty}^{\infty} \hspace{-3.5mm} \alpha_{(k_1,l_1)}(t) \, \alpha^*_{(k_2,l_2)}(t) \, dt & \hspace{-4mm} = & \hspace{-4mm} \frac{T}{MN} \sum\limits_{n_1 = 0}^{N-1} \sum\limits_{n_2 = 0}^{N-1} \hspace{-1mm} {\Bigg [} e^{j 2 \pi \frac{n_1 k_1 - n_2 k_2}{N} } \nonumber \\
& & \hspace{-7mm} (M \Delta f)^2 e^{j \pi M \Delta f \left(\frac{(l_2 - l_1)T}{M} + (n_2 - n_1)T  \right)}  \nonumber \\
& & \hspace{-43mm} \int_{-\infty}^{\infty} \hspace{-4mm}  \mbox{\small{sinc}} \hspace{-1mm} \left( \hspace{-1mm} M \Delta f  \hspace{-1mm} \left(\hspace{-1mm} t - \frac{l_1T}{M} - n_1 T \right)\hspace{-1mm} \right) \mbox{\small{sinc}} \hspace{-1mm} \left(M \Delta f \hspace{-1mm} \left( \hspace{-1mm} t - \frac{l_2T}{M} - n_2 T \right) \hspace{-1mm} \right) dt {\Bigg ]} \nonumber \\
& \hspace{-70mm} \mya & \hspace{-36mm} \frac{T}{MN} \sum\limits_{n_1 = 0}^{N-1} \sum\limits_{n_2 = 0}^{N-1} \hspace{-1mm} {\Bigg [} e^{j 2 \pi \frac{n_1 k_1 - n_2 k_2}{N} } e^{j \pi M \Delta f \left(\frac{(l_2 - l_1)T}{M} + (n_2 - n_1)T  \right)} \nonumber \\
& & \hspace{-32mm}  (M \Delta f) \mbox{\small{sinc}}\left(M \Delta f \left(\frac{(l_2 - l_1)T}{M} + (n_2 - n_1) T \right) \right) {\Bigg ]} \nonumber \\
& \hspace{-70mm} \myb & \hspace{-36mm} \frac{1}{N} \sum\limits_{n_1 = 0}^{N-1} \sum\limits_{n_2 = 0}^{N-1} \hspace{-1mm} {\Bigg [} e^{j 2 \pi \frac{n_1 k_1 - n_2 k_2}{N} } e^{j \pi M \Delta f \left(\frac{(l_2 - l_1)T}{M} + (n_2 - n_1)T  \right)} \nonumber \\
& & \delta[l_1 - l_2] \delta[n_1 - n_2] {\Bigg ]} \nonumber \\
& \hspace{-72mm} = & \hspace{-38mm} \delta[l_1 - l_2]  \left(  \frac{1}{N} \sum\limits_{n_1=0}^{N-1} e^{j 2 \pi \frac{n_1 (k_1 - k_2)}{N}} \right) \, = \, \delta[l_1 - l_2] \delta[k_1 - k_2]
\end{eqnarray} 
\normalsize}
where step (a) follows from the fact that $\int_{-\infty}^{\infty} W^2 \mbox{\small{sinc}}(W(t - \tau_1)) \mbox{\small{sinc}}(W(t - \tau_2)) dt = W \mbox{\small{sinc}}(W(\tau_2 - \tau_1))$.
Step (b) follows from the fact that $\mbox{\small{sinc}}\left(M \Delta f \left(\frac{(l_2 - l_1)T}{M} + (n_2 - n_1) T \right) \right) =  \mbox{\small{sinc}}\left( l_2 - l_1 + M(n_2 - n_1)\right)$ which is
one only when $l_1 = l_2$ and $n_1 = n_2$, and is otherwise zero. This is because, $l_1,l_2 =0,1,\cdots, M-1$ and therefore $(l_2 - l_1)$ can never be a non-zero integer multiple of $M$. This shows that $\boldsymbol \alpha$ is a basis with $MN$ orthonormal signals and is therefore
$MN$-dimensional.

\section{Proof of Theorem \ref{thm73}}
\label{prfthm73}
The ZAK representation of $y(t)$ is given by

{\vspace{-4mm}
\small
\begin{eqnarray}
\label{eqn57}
\hspace{-1mm} {\mathcal Z}_y(\tau, \nu) & \hspace{-3mm}  \mya & \hspace{-3mm} \sqrt{T} \hspace{-1mm} \sum\limits_{n'=0}^{N} \hspace{-1mm} y(\tau + n' T) e^{-j 2 \pi \nu n' T}, \,\, 0 \leq \tau < T \,,\, 0 \leq \nu < \Delta f \nonumber \\
& &  \hspace{-15mm} \myb \,  \sum\limits_{i=1}^{L} h_i \, e^{j 2 \pi \nu_i (\tau - \tau_i)} \, {\mathcal Z}_x(\tau - \tau_i, \nu - \nu_i) \,\, + \,\, {\mathcal Z}_n(\tau, \nu)
\end{eqnarray}
\normalsize}
where step (a) follows from (\ref{zdef1}) and step (b) follows from Result \ref{zlem52}. Here ${\mathcal Z}_x(\tau, \nu)$ and ${\mathcal Z}_n(\tau, \nu)$ are the ZAK representations of $x(t)$ and $n(t)$ respectively.
From (\ref{eqn44}), the ZAK representation of $x(t)$ is given by
\begin{eqnarray}
\label{eqn58}
{\mathcal Z}_x(\tau, \nu)  & \hspace{-3mm}  \mya &  \hspace{-3mm} \frac{1}{\sqrt{MN}} \sum\limits_{k=0}^{N-1} \sum\limits_{l=0}^{M-1} x[k,l] {\mathcal Z}_{\psi, \frac{lT}{M}, \frac{k \Delta f}{N}}(\tau, \nu)
\end{eqnarray}
where ${\mathcal Z}_{\psi,\frac{lT}{M}, \frac{k \Delta f}{N} }(\tau, \nu)$ is the ZAK representation of  $\psi^{(q,s)}_{( lT/M, k /NT)}(t)$ (see (\ref{eqn43})).
Step (a) follows from the linearity of the ZAK representation (i.e., the ZAK representation of the sum of two TD signals is
the sum of their ZAK representations), and the fact that $\alpha_{(k,l)}(t) = \frac{1}{\sqrt{MN}} \psi^{(q,s)}_{( lT/M, k /NT)}(t)$ (see (\ref{eqn27})).
From Theorem \ref{thm21}, the ZAK representation of $\psi^{(q,s)}_{( lT/M, k /NT)}(t)$ is given by

{\vspace{-4mm}
\small
\begin{eqnarray}
\label{eqn52}
{\mathcal Z}_{\psi, \frac{lT}{M}, \frac{k \Delta f}{N}}(\tau, \nu) & \hspace{-3mm} = &  \hspace{-3mm}  {\mathcal Z}_q\left( \frac{lT}{M} , \nu - \frac{k \Delta f}{N} \right) \, {\mathcal Z}_s\left( \tau - \frac{lT}{M}, \nu \right)  \nonumber \\
& & \hspace{-32mm} \mya   e^{-j \pi \left( \nu  - \frac{k \Delta f}{N} \right) (N-1) T} \, \frac{\sin\left( \pi  \left( \nu  - \frac{k \Delta f}{N} \right) N T \right)}{\sin\left( \pi \left( \nu  - \frac{k \Delta f}{N} \right) T \right)} e^{j 2 \pi \nu \left( \tau - \frac{lT}{M} \right)} \nonumber \\
& & \hspace{-32mm}    e^{-j 2 \pi \left\lfloor \frac{\nu}{\Delta f}  \right\rfloor  \Delta f  \left( \tau - \frac{lT}{M}  \right) } e^{j \pi (M-1) \Delta f \left( \tau - \frac{lT}{M} \right)}  \frac{\sin( \pi M \Delta f \left( \tau - \frac{lT}{M}  \right)  )}{\sin(\pi \Delta f \left( \tau - \frac{lT}{M}  \right) )}
\end{eqnarray}
\normalsize}
where step (a) follows from the expressions of ${\mathcal Z}_q(\tau, \nu)$ and ${\mathcal Z}_s(\tau, \nu)$ in (\ref{expreqn1}).
Using (\ref{eqn52}) in (\ref{eqn58}) we get an expression for ${\mathcal Z}_x(\tau, \nu)$ in terms of the information symbols $x[k,l], k=0,1,\cdots, N-1, l=0,1,\cdots, M-1$.
Using this expression in (\ref{eqn57}) we then get an expression for ${\mathcal Z}_y(\tau, \nu)$ in terms of the information symbols $x[k,l]$. Using this expression, the sampled DD domain
signal $Y[k',l'] = {\mathcal Z}_y\left( \tau = \frac{l'T}{M} , \nu = \frac{k' \Delta f}{N}  \right)$ is given by (\ref{eqn67}).

\section{Proof of Theorem \ref{thm8}}
\label{prfthm8}
In (\ref{eqn67}), $Z[k',l']$ is given by
\begin{eqnarray}
\label{eqn79}
Z[k',l'] & = &  {\mathcal Z}_n\left(\tau = \frac{l'T}{M} , \nu = \frac{k' \Delta f}{N} \right) \nonumber \\
& \mya & \sqrt{T} \sum\limits_{n'=0}^{N} n\left(n'T + \frac{l'T}{M}  \right) \, e^{- j 2 \pi \frac{n' k'}{N} }
\end{eqnarray} 
where step (a) follows from the fact that the received TD signal $y(t)$ is limited to the interval $[0 \,,\, (N+1)T )$.
Let the information symbols $x[k,l] \, \sim \, \mbox{\small{i.i.d.}} \, {\mathcal C}{\mathcal N}(0 , \rho)$. Then, since the signals $\alpha_{(k,l)}(t)$ belong to the orthonormal basis $\boldsymbol \alpha$ (see Theorem \ref{lem18}), from
the first equation in (\ref{eqn44}) it follows that
\begin{eqnarray}
{\mathbb E} \left[ \int_{-\infty}^{\infty} \left\vert x(t) \right\vert^2 \, dt  \right] & = & MN \rho.
\end{eqnarray}
Since the transmit signal $x(t)$ has most of its energy in the time-interval $[0 \,,\, NT)$, the average 
transmit power is $MN \rho/NT = M \rho/T$. Let the power spectral density (PSD) of AWGN be unity. Since the communication bandwidth is $M \Delta f$, the AWGN power at the
receiver is $M\Delta f$. Therefore, the ratio of the transmit power to the receiver noise power is
\begin{eqnarray}
\frac{M \rho / T}{M \Delta f } \, = \, \rho.
\end{eqnarray}
Let the sampled DD domain signal $Y[k',l']= {\mathcal Z}_y\left(\tau = \frac{l'T}{M}, \nu = \frac{k' \Delta f}{N} \right), k'=0,1,\cdots, N-1, l'=0,1,\cdots, M-1$ be organized into a vector ${{\bf  y}} \in {\mathbb C}^{MN \times 1}$, where
the $(k'M + l' + 1)$-th element of  ${{\bf y}}$ is $Y[k',l']$, $k'=0,1,\cdots,N-1, l'=0,1,\cdots, M-1$. Similarly, $x[k,l]$ is organized into the information symbol vector ${\bf x} \in {\mathbb C}^{MN \times 1}$
where the $(kM+l+1)$-th element of ${\bf x}$ is $x[k,l]$. Also, let ${\Tilde {\bf H}} \in {\mathbb C}^{MN \times MN}$ be the effective DD domain channel matrix whose element in its
$(k'M + l' +1)$-th row and $(kM + l+1)$-th column is ${\Tilde h}[k',l',k,l]$ (${\Tilde h}[k',l',k,l]$ is defined in (\ref{eqn67})).
From (\ref{eqn67}) it then follows that
\begin{eqnarray}
\label{eqn71}
{ {\bf y}} & = &  \sqrt{MN} \, {\Tilde {\bf H}} \, {\bf x} \, + \, {\bf z}
\end{eqnarray}
where ${\bf z} \in {\mathbb C}^{MN \times 1}$ is the vector of noise samples, i.e.,
the $(k'M + l' + 1)$-th element of ${\bf z}$ is $Z[k',l']$. Since the communication bandwidth is $M \Delta f$ and the PSD of AWGN is unity,
the noise samples $n\left(n'T + \frac{l'T}{M}  \right), \, n'=0,1,\cdots, N, l'=0,1,\cdots, M-1$ are i.i.d. ${\mathcal C}{\mathcal N}(0, M \Delta f )$.
The covariance matrix of the noise vector ${\bf z}$ is denoted by ${\bf K}_{\bf z}$. The expression for ${\bf K}_{\bf z}$ follows from equation $(22)$ in
\cite{Saif2020}, i.e., the element of ${\bf K}_{\bf z}$ in its $(k'M + l' +1)$-th row and $(kM + l +1)$-th column is given by

{\vspace{-4mm}
\small
\begin{eqnarray}
\label{eqn81}
{K}_{\bf z}[{k'M + l' +1,kM + l +1}]  & \hspace{-3mm} =  & \hspace{-3mm}  {\mathbb E} \left[  Z[k',l'] Z^*[k,l]   \right]\nonumber \\
& \hspace{-12mm} \mya  & \hspace{-7mm}
\begin{cases}
 M N \left(1 + \frac{1}{N} \right) , k' = k \,,\, l' = l \\
 M N \left( \frac{1}{N} \right) \hspace{6mm} , k' \ne k \,,\, l' = l \\
0  \hspace{11mm} \,, l' \ne l \\
\end{cases}, \nonumber \\
& & \hspace{-32mm} k',k = 0,1,\cdots, N-1 \,,\, l',l = 0,1,\cdots, M-1.
\end{eqnarray}
\normalsize}
where step (a) can be derived from the expression of $Z[k',l']$ in the R.H.S. of (\ref{eqn79}). 
From (\ref{eqn71}) it then follows that the spectral efficiency (SE) achieved by the DD domain
modulation in (\ref{eqn44}) is given by \cite{DTse}
\begin{eqnarray}
C & = &  \frac{1}{MN} \, \log_2 \left\vert  {\bf I} \,  + \, M N \rho  \, {\Tilde {\bf H}}^H  {{\bf K}_{{\bf z}}}^{-1}  {\Tilde {\bf H}}  \right\vert \nonumber \\
& = & \frac{1}{MN} \, \log_2 \left\vert  {\bf I} \,  + \, \rho  \, {\Tilde {\bf H}}^H  {\Tilde {\bf K}}^{-1}  {\Tilde {\bf H}}  \right\vert,  \nonumber \\
{\Tilde {\bf K}} & \Define & \frac{1}{MN} {{\bf K}_{{\bf z}}}
\end{eqnarray}
where the elements of ${{\bf K}_{{\bf z}}}$ are given by (\ref{eqn81}).

\section{Fraction of Interfered Information Symbols in CP-OFDM}
\label{cpofdm}
In OFDM, the $M$ orthonormal basis
signals are

{\vspace{-4mm}
\small
\begin{eqnarray}
\label{ofdmeqn1}
\phi_{k}(t) & = \begin{cases}
\frac{1}{\sqrt{T}} e^{j 2 \pi k \frac{t}{T}} \,&,\,  0 \leq t < T \\
0 \, &, \, \mbox{\small{otherwise}}
\end{cases} \nonumber \\
& & \hspace{-50mm} k=0,1,\cdots, M-1.
\end{eqnarray}
\normalsize}
With a cyclic prefix equal to the path delay $\tau'$, and information symbols $x[k], k=0,1,\cdots, M-1$,
the transmit CP-OFDM signal is

{\vspace{-4mm}
\small
\begin{eqnarray}
\label{ofdmeqn2}
x_{\mbox{\tiny{ofdm}}}(t) & = \begin{cases} 
\sum\limits_{k=0}^{M-1} x[k] \, \phi_k(t)  \, &, \, 0 \leq t < T \\
\sum\limits_{k=0}^{M-1} x[k] \, \phi_k(t+T)   \,  &, \, - \tau' \leq t < 0 \\
0 \, &, \, \mbox{\small{otherwise}}
\end{cases}.
\end{eqnarray}
\normalsize}
The received noise-free TD signal is given by
\begin{eqnarray}
\label{ofdmeqn3}
y_{\mbox{\tiny{ofdm}}}(t)  & = &  h' \, e^{j 2 \pi \nu' (t - \tau')}  \, x_{\mbox{\tiny{ofdm}}}(t - \tau') 
\end{eqnarray}
where $h'$ is the channel gain of the single channel path.
The receiver removes the CP and computes

{\vspace{-4mm}
\small
\begin{eqnarray}
\label{ofdmeqn4}
Y_{\mbox{\tiny{ofdm}}}[m] & \hspace{-3mm}  \Define &  \hspace{-3mm} \int_{0}^{T}  \hspace{-2mm} y_{\mbox{\tiny{ofdm}}}(t)  \,  \phi_{m}^*(t) \, dt \nonumber \\
 & \hspace{-3mm}   \mya &   \hspace{-3mm}  \sum\limits_{k=0}^{M-1}  x[k] \, H_{\mbox{\tiny{ofdm}}}[m,k],  \,\, m=0,1,\cdots, M-1, \nonumber \\
H_{\mbox{\tiny{ofdm}}}[m,k]  & \Define &  h' \, \int_{0}^T \hspace{-1mm} e^{j 2 \pi \nu' (t - \tau')} \, \phi_k(t - \tau') \, \phi_m^*(t) \, dt  \nonumber \\
& & \hspace{-22mm} = h' e^{-j 2 \pi \frac{\tau'}{T}  \left({\nu' T}  + k \right)} \, e^{j \pi \left(  {\nu' T}  + k - m \right)}  \, \mbox{\small{sinc}}\left(  {\nu' T} + k - m \right)
\end{eqnarray}
\normalsize}
where step (a) follows from substituting the R.H.S. of (\ref{ofdmeqn2}) in (\ref{ofdmeqn3}), and then using the resulting expression
for $y_{\mbox{\tiny{ofdm}}}(t)$ in the integral of the first equation in (\ref{ofdmeqn4}).
From (\ref{ofdmeqn4}) it is clear that due to the channel induced Doppler shift $\nu'$, an information symbol $x[k]$ transmitted on the $k$-th sub-carrier is
received as $H[m,k] x[k]$ on the $m$-th sub-carrier.
Since $\vert H_{\mbox{\tiny{ofdm}}}[m,k] \vert^2 = \vert h' \vert^2 \, \mbox{\small{sinc}}^2(\nu' T + k - m )$ (see (\ref{ofdmeqn4})), it follows that most of the energy of $x[k]$ is received in and around the $m=  \left\lfloor k + {\nu' T}  \right\rfloor$-th
sub-carrier. Next, for a given $k \in \{ 0,1,\cdots, M-1 \}$, let ${\mathcal G}_k$ denote the smallest cardinality set of sub-carrier indices such that
the fraction of total energy of $x[k]$ received in the sub-carrier indices in ${\mathcal G}_k$ is at least $0.99$, i.e.,

{\vspace{-4mm}
\small
\begin{eqnarray}
{\mathcal G}_k & \hspace{-3mm}  \Define &  \hspace{-3mm} \arg \hspace{-9mm} \min_{{\mathcal D} \subseteq {\mathcal V}  \,  {\Big \vert}  \,  \frac{\sum\limits_{m \in {\mathcal D} } \left\vert  H_{\mbox{\tiny{ofdm}}}[m,k]  \right\vert^2}{\sum\limits_{m \in {\mathcal V}} \left\vert  H_{\mbox{\tiny{ofdm}}}[m,k]  \right\vert^2 } \geq 0.99}  \hspace{-8mm}  \left\vert  {\mathcal D}  \right\vert 
 \,\,\,\,\,  =   \,\,\,\,\,  \arg \hspace{-9mm} \min_{{\mathcal D} \subseteq {\mathcal V} \,  {\Big \vert}  \,  \frac{\sum\limits_{m \in {\mathcal D} }  \mbox{\tiny{sinc}}^2\left(  {\nu'T} + k - m \right)}{\sum\limits_{m \in {\mathcal V}}  \mbox{\tiny{sinc}}^2\left(  {\nu'T} + k - m \right)} \geq 0.99}  \hspace{-8mm}  \left\vert  {\mathcal D}  \right\vert, \nonumber \\
{\mathcal V} & \Define & \{ 0, 1, \cdots, M-1 \}.
\end{eqnarray}
\normalsize}
The fraction of information symbols which are interfered by $x[k]$ is then given by

{\vspace{-4mm}
\small
\begin{eqnarray}
\mbox{\small{Frac. of interfered symbols}} & = & \frac{\left\vert {\mathcal G}_k \right\vert - 1}{M-1}.
\end{eqnarray}
\normalsize}

\end{document}